\documentclass[12pt,preprint]{aastex}

\newcommand{\bzcat}{ROMA-BZCAT}

\newcommand{\chn}{{\it Chandra}}

\newcommand{\fer}{{\it Fermi}}

\newcommand{\swf}{{\it Swift}}
\newcommand{\suz}{{\it Suzaku}}
\newcommand{\xmm}{{\it XMM-Newton}}
\newcommand{\wse}{{\it WISE}}

\usepackage{graphicx}
\usepackage{longtable}

\usepackage{lineno}
\linenumbers

\slugcomment{version \today: fm}
\shorttitle{Unidentified Gamma-ray Sources II}
\shortauthors{F. Massaro et al. 2013}

\begin{document}
\title{Unveiling the nature of the unidentified gamma-ray sources II: \\ radio, infrared and optical counterparts of the gamma-ray blazar candidates}
\author{
F. Massaro\altaffilmark{1}, 
R. D'Abrusco\altaffilmark{2}, 
A. Paggi\altaffilmark{2}, 
N. Masetti\altaffilmark{3},
M. Giroletti\altaffilmark{4}, \\
G. Tosti\altaffilmark{5,6}, 
Howard A. Smith\altaffilmark{2}, 
\& 
S. Funk\altaffilmark{1}.}

\altaffiltext{1}{SLAC National Laboratory and Kavli Institute for Particle Astrophysics and Cosmology, 2575 Sand Hill Road, Menlo Park, CA 94025, USA}
\altaffiltext{2}{Harvard - Smithsonian Astrophysical Observatory, 60 Garden Street, Cambridge, MA 02138, USA}
\altaffiltext{3}{INAF - Istituto di Astrofisica Spaziale e Fisica Cosmica di Bologna, via Gobetti 101, 40129, Bologna, Italy}
\altaffiltext{4}{INAF Istituto di Radioastronomia, via Gobetti 101, 40129, Bologna, Italy}
\altaffiltext{5}{Dipartimento di Fisica, Universit\`a degli Studi di Perugia, 06123 Perugia, Italy}
\altaffiltext{6}{Istituto Nazionale di Fisica Nucleare, Sezione di Perugia, 06123 Perugia, Italy}

\begin{abstract}
A significant fraction ($\sim$30\%) of the high-energy gamma-ray sources listed in the 
second \fer\ LAT catalog (2FGL) are still of unknown origin, 
being not yet associated with counterparts at low energies.
We recently developed a new association method to identify
if there is a $\gamma$-ray blazar candidate within the positional 
uncertainty region of a generic 2FGL source.
This method is entirely based on the discovery that blazars 
have distinct infrared colors with respect to other extragalactic sources
found thanks, to the Wide-field Infrared Survey Explorer (\wse) all-sky observations.
Several improvements have been also performed to increase the efficiency of our method
in recognizing $\gamma$-ray blazar candidates.
In this paper we applied our method to two different samples, the first constituted by the 
unidentified $\gamma$-ray sources (UGSs) while the second by the
active galaxies of uncertain type (AGUs), both listed in the 2FGL. 
We present a catalog of IR counterparts for $\sim$20\% of the UGSs investigated.
Then, we also compare our results on the associated sources with those present in literature. 
In addition, we illustrate the extensive archival research carried out to identify 
the radio, infrared, optical and X-ray counterparts 
of the \wse\ selected, $\gamma$-ray blazar candidates.
Finally, we discuss the future developments of our method 
based on ground-based { follow-up} observations.
\end{abstract}


\section{Introduction}
\label{sec:intro}
Unveiling the nature of the Unidentified Gamma-ray Sources (UGSs) \citep[e.g.,][]{abdo09}
is one of the biggest challenges in contemporary gamma-ray astronomy.
Since the era of the Compton Gamma-ray Observatory 
{ many} $\gamma$-ray objects have { not} been conclusively associated 
with counterparts at other frequencies \citep{hartman99}, 
although various classes have been investigated
to understand whether they are likely to be detected at $\gamma$-ray energies or not
\citep[e.g.,][]{thompson08}. 

According to the {Second \fer\ Large Area Telescope (LAT) catalog} \citep[2FGL;][]{nolan12}, 
$\sim$1/3 of the $\gamma$-ray detected sources are still unassociated with their low energy counterparts. 
Moreover a large fraction of the UGSs { are likely to be} of blazars, the rarest class of radio loud active galactic nuclei, because
their emission dominates the $\gamma$-ray sky \citep[e.g.,][]{mukherjee97,abdo10}.
However, due to the incompleteness of the current radio and X-ray surveys on the basis of the { usual}
$\gamma$-ray association method is not always possible to find the blazar-like counterpart of an UGS.
Additional attempts have { also} been recently developed to associate or to characterize the UGSs 
using { either} pointed \swf\ observations \citep[e.g.,][]{mirabal09a,mirabal09b} 
or statistical approaches \citep[e.g.][]{mirabal10,ackermann12}. 

Blazar emission is characterized by high and variable polarization, 
apparent superluminal motions, and high luminosities,
generally combined with a flat radio spectrum that steepens toward the infrared-optical bands and
together with rapid flux variability from the radio to $\gamma$-rays \citep[e.g.,][]{urry95}.
Their spectral energy distributions show two main 
broad components: a low-energy one peaking in the range from the IR to the X-ray band, 
and a high-energy component peaking from MeV to TeV energies \citep[e.g.,][]{giommi05}.

Blazars are divided in two main classes: the low luminosity { class} constituted by the BL Lac objects
and characterized by featureless optical spectra, and
the second { class} composed of flat-spectrum radio quasars 
that show optical emission lines, typical of quasar spectra \citep{stickel91,stoke91}.
In the following we label the BL Lac objects as BZBs and the 
flat-spectrum radio quasars as BZQs, following the nomenclature 
of the { Multifrequency Catalogue of Blazars} \citep[\bzcat,][]{massaro09,massaro10,massaro11}.

On the basis of the preliminary data release of the Wide-field Infrared Survey Explorer
\citep[\wse, see][for more details]{wright10}\footnote{http://wise2.ipac.caltech.edu/docs/release/prelim/}, 
we discovered that in the 3-dimensional IR color space $\gamma$-ray emitting blazars
lie in a distinct region, well separated from other extragalactic 
sources { whose} IR emission is dominated by thermal radiation 
\citep[e.g.,][]{paper1,paper2}.

According to D'Abrusco et al. (2013) we refer to the 3-dimensional region occupied 
by $\gamma$-ray emitting blazars as the $locus$, 
{  to its  2-dimensional projection in the [3.4]-[4.6]-[12] $\mu$m color-color diagram as the \wse\ Gamma-ray Strip.}

{ This \wse\ analysis} led to the development of a new association 
method to recognize $\gamma$-ray blazar candidates
{ for the} unidentified $\gamma$-ray sources { listed} in the 2FGL
\citep{paper3,paper4}, as well as in the 4$^{th}$ 
{\it INTEGRAL} catalog \citep{paper5}.

{ In the present paper we adopt} several improvements { recently made on} the association procedure 
and { we use} a more conservative approach \citep[see][for more details]{paper6},
mostly based on the \wse\ full archive
\footnote{http://wise2.ipac.caltech.edu/docs/release/allsky/}, available since March 2012 \citep[see also][]{cutri12}.
{ We successfully tested} the association procedure on all the blazars listed
in the Second \fer\ LAT Catalog of active galactic nuclei \citep[2LAC;][]{ackermann11}
and in the 2FGL catalogs, to estimate its efficiency and its completeness.
 
In this paper we apply this method to the UGSs and to
sample of the active galactic nuclei of uncertain type (AGUs) that have still unclear classification
(see 2FGL and also Section~\ref{sec:agu} for specific definition of the class), both listed in the 2FGL.
We also performed an extensive literature search looking for 
multifrequency information { on} the $\gamma$-ray blazar candidates selected on the basis of their \wse\ colors
to confirm their nature. 
{ As we show below} this research is crucial { to determine whether or not there are classes}  
of Galactic and extragalactic sources that, having IR colors similar to those of blazars,
could be a contaminants of the association method.

The paper is organized as follows: in Section~\ref{sec:sample} 
we describe the sample selected.
In Section~\ref{sec:method} we illustrate the basic details of the association procedure 
{ and} highlight the improvements with respect to the previous version.
In Section~\ref{sec:results} we describe the results obtained.
Section~\ref{sec:counterparts} is dedicated to the 
{ correlating our results with} several databases at radio, infrared, optical
and X-ray frequencies to characterize the multifrequency behavior of the 
$\gamma$-ray blazar candidates.
{ We then} compare our results on the associated sources with those based on 
statistical methods developed by Ackermann et al. (2012) in Section~\ref{sec:comparison}. 
Finally, Section~\ref{sec:summary} is devoted to our conclusions. 

{ The most frequent acronyms used in the paper are listed in Table~\ref{tab:acronym}.}
\begin{table}
\caption{List of most frequent acronyms.}
\begin{tabular}{|lc|}
\hline
Name & Acronym \\
\hline
\noalign{\smallskip}
Multifrequency Catalog of blazars & \bzcat\ \\ 
First \fer\ Large Area Telescope catalog & 1FGL \\
Second \fer\ Large Area Telescope Catalog & 2FGL \\
Second \fer\ LAT Catalog of Active Galaxies & 2LAC \\
\hline
\noalign{\smallskip}
BL Lac object & BZB \\
Flat Spectrum Radio Quasar & BZQ \\
Blazar of Uncertain type & BZU \\
\hline
\noalign{\smallskip}
Unidentified Gamma-ray Source & UGS \\
Active Galactic nucleus of Uncertain type & AGUs \\
\noalign{\smallskip}
\hline
\end{tabular}\\
\label{tab:acronym}
\end{table}

\section{Sample selection}
\label{sec:sample}

\subsection{The unidentified gamma-ray sources}
\label{sec:ugs}
{ Our primary sample of UGSs consists of all the sources for which no counterpart was assigned at low energies
in the 2FGL or in the 2LAC \citep[][respectively]{nolan12,ackermann11}, for a total of 590 $\gamma$-ray objects.

We considered and analyzed independently two subsamples of UGSs, distinguishing the 299 \fer\ sources
without any $\gamma$-ray analysis flags from the other 291 objects
that have a warning in their $\gamma$-ray detection.
This distinction has been performed because future releases of the \fer\ catalogs
based on improvements of the \fer\ response matrices and revised analyses,}
could make their detection more reliable, as occurred for a handful of sources
flagged in the first \fer\ LAT catalog \citep[1FGL,][]{abdo10,nolan12}.

\subsection{The active galaxies of uncertain type}
\label{sec:agu}
According to the definition of the 2LAC and 2FGL { catalogs},
active galaxies of uncertain type (AGUs) 
are $\gamma$-ray emitting sources { with} at least one of the following criteria:
\begin{enumerate}
\item they do not have a good optical spectrum available or with an uncertain classification, as for example, sources classified as { blazars of uncertain type (BZU)} in the \bzcat;
\item they have been selected as candidate counterparts
on the basis of the $log N$ - $log S$ and the Likelihood Ratio methods described in the 2LAC and applied to several radio catalogs: { including} the 
AT20G \citep{murphy10}, CRATES \citep{healey07}, or CLASS \citep{falco98} \citep[see][for details]{ackermann11};
\item they are coincident with a radio and a X-ray source selected by the Likelihood Ratio method.
\end{enumerate}

The number of AGUs in the 2FGL, { that have been analyzed} is 210;
{ excluding} $\gamma$-ray sources { with} analysis flags 
(defined according to both the 2FGL or the 2LAC descriptions).

\section{The Association Procedure}
\label{sec:method}
The complete description of our association procedure together with the estimates of its efficiency and its completeness 
can be found in D'Abrusco et al. (2013) where we discuss a new and improved version 
of the association method based on a 3-dimensional parametrization of the $locus$ occupied by  
$\gamma$-ray emitting blazars with \wse\ counterparts. Here we provide only an overview.
We note that the results of the improved method are in agreement with those of the previous 
parametrization, thus superseding the previous procedure \citep{paper1,paper3,paper4}.

The new association procedure { was} built to improve the efficiency of recognizing $\gamma$-ray blazar candidates,
to decrease the number of possible contaminants and, at the same time, to determine if { a} selected $\gamma$-ray blazar counterpart
is more likely to be a BZB or a BZQ.
The main differences between the two association methods reside in the parameter space where the $locus$ has been defined
(IR color space for the old version and principal component space for the new one) and in the 
assignment criteria of the classes for the $\gamma$-ray blazar candidates \citep[see][]{paper6}.
The new method { also} takes into account of the correction for Galactic extinction
for all the \wse\ magnitudes\footnote{The IR magnitudes in the [3.4], [4.6], [12], [22] $\mu$m nominal \wse\ bands 
are in the Vega system.} according to the Draine (2003) relation.
As shown in D'Abrusco et al. (2013), this correction affects only marginally the [3.4]-[4.6] color,
in particular at low Galactic latitudes (i.e., $|b|<$15 deg).

{ The principal component analysis is designed to reduce the dimensionality of a dataset consisting of usually 
large number of correlated variables while retaining as much as possible of the variance present in the data 
in the smallest possible number of orthogonal parameters. 
This is achieved by transforming the observed parameter into a new set of variables, the principal components. 
They are ordered so that the first accounts the largest possible variance of the original dataset 
and the others in turn have the highest variance possible under the constraint of being orthogonal to the preceding ones \citep[e.g.,][]{pearson01,jolliffe02}.
Thus our new parametrization of the $locus$ in the PC space, where the maximum variance is contained 
along only one axis, is simpler than any other possible representation in the IR color space.}

For each $\gamma$-ray source we defined a {\it search region}:
a circular region of radius $\theta_{95}$ equal to the semi-major axis of the ellipse 
corresponding to the positional uncertainty region of the \fer\ source at 95\% level of confidence { and} 
centered at the 2FGL position of the $\gamma$-ray source \citep[e.g., ][]{nolan12}.
We selected and calculated the IR colors for the \wse\ sources within the 
{\it search region} detected in all four bands.

To compare the infrared colors of generic infrared sources that lie in the {\it search region}
with those of the $\gamma$-ray emitting ones, we developed a 3-dimensional parametrization of the $locus$
in the parameter space of its principal components.
The $locus$ was described as a cylinder in the space of the principal components.
This choice simplifies and improves the previous description 
built using irregular quadrilaterals on all the color-color diagrams \citep{paper3}. 
{ Moreover, the cylinder axis is aligned along the first PC axis, 
which accounts for the larger fraction possible of the variance of the dataset in the IR color space, is
the simplest parametrization available.}

We then assign to each source {\it score} value $s$
that is a proxy of the distance between the $locus$ { surface}
and the source location in the 3-dimensional parameter space of the principal components.
{ The values of $s$ allow} to to evaluate if the IR colors of a generic source are consistent with those of the known $\gamma$-ray emitting blazars.
{ They were} weighted taking into account of all the color errors and they are also normalized between 0 and 1.
We define three classes (i.e., A, B, C) of reliability for the $\gamma$-ray blazar candidates.
A generic source is assigned to class A, class B or class C when its $score$
his higher than the threshold values defined by the 90\%, 60\% and 30\% percentiles
of the $score$ distributions of all the $\gamma$-ray blazars that constitute the $locus$, respectively.
{ We consider reliable $\gamma$-ray blazar candidates
only those having the score higher than 70\% of their distributions.}
Thus sources with high values of the $score$ (e.g., $>$0.8) are very likely to be blazars
and belong to class A, while sources with $score$ values $\sim$0.5 belong to class C and are less probable $\gamma$-ray blazars. 
IR sources that having $score$ values null or extremely low (e.g., $\sim$0.1) were marked as {\it outliers} and were not 
considered as $\gamma$-ray blazar candidates \citep[see][for an extensive explanation on the class definitions]{paper6}.

The $locus$ { was} divided in subregions on the basis of the space density of BZBs and BZQs
in the parameter space of its principal components, { }thereby permitting us to determine if a selected $\gamma$-ray blazar candidate is
more likely { to be} a BZB or a BZQ.

{ Finally, we ranked all the \wse\ sources within each {\it search region} and selected as best candidate counterpart for the UGS
the one with the highest class; when more than one candidate of the same class was present, we chose the one closest to the
$\gamma$-ray position as best one.}

\section{Results}
\label{sec:results}

\subsection{The unidentified gamma-ray sources}
\label{sec:resugs}
For the UGSs without $\gamma$-ray analysis flags
we found 75 $\gamma$-ray blazar candidates out of the 299 objects analyzed:
8 sources { have}  2 candidates, 1 source { has}  3, and 1 source { has} 4 candidates,
{ while} 52 associations are unique.
We found 2 $\gamma$-ray blazar candidates of class A, 12 of class B and 61 of class C, respectively, 
in the whole sample of 75 sources; 32 of them are classified as BZB type, 
29 as BZQ type and the remaining 14 are still uncertain \citep[see][for more details]{paper6}.
{ All our $\gamma$-ray blazar candidates have a {\it signal-to-noise ratio} systematically larger than 10.9 in the \wse\ band centered at 12$\mu$m
and larger than $\sim$20 for the 3.4$\mu$m and 4.6$\mu$m nominal bands.}
For all these 75 sources we performed a cross correlation with the major radio, infrared, optical,
and X-ray surveys (see Section~\ref{sec:counterparts}).

In the sample of UGSs with $\gamma$-ray analysis flags
we found 71 $\gamma$-ray blazar candidates out of the 291 objects investigated:
6 sources { have} 2 candidates, 4 sources have 3 candidates, 2 sources have 4 and 6 candidates, respectively,
{ while} 35 associations are unique.
We found 8 $\gamma$-ray blazar candidates of class A, 20 of class B and 43 of class C, respectively, 
in the whole sample of 71 sources; 36 of them are classified as BZB type, 
22 as BZQ type and the remaining 13 are still uncertain \citep[see][]{paper6}.
{ We also performed the cross correlation with the major radio, infrared, optical,
and X-ray databes for these 71 UGSs listed in the Section~\ref{sec:counterparts}.}

\subsection{The active galaxies of uncertain type}
\label{sec:resagu}
For the AGU sample we found 125 $\gamma$-ray blazar candidates out of the 210 sources analyzed:
10 sources { have} 2 candidates within their {\it search region}, while
the remaining 105 candidates { have} unique associations.
There are 10 $\gamma$-ray blazar candidates of class A, 39 of class B and 76 of class C, respectively, 
in the whole sample of 125 sources; 52 out of 125 are classified as BZB type
on the basis of the IR colors of blazars of similar type, 
39 as BZQ type and the remaining 34 are still uncertain \citep[see][for more details]{paper6}.
Eighty-seven sources out of 125 associations correspond to those reported in the 2LAC or in the 2FGL.
{ All our $\gamma$-ray blazar candidates have a {\it signal-to-noise ratio} systematically larger than 10.9 in the \wse\ band centered at 12$\mu$m
and larger than $\sim$20 for the 3.4$\mu$m and 4.6$\mu$m nominal bands.}
In these case we did not provide any additional radio or X-ray information since it is already 
present in both the 2LAC and the 2FGL, 
while a multifrequency investigation has been performed for the remaining 38.
Additional IR information for all the AGUs associated 
will be discussed in Section~\ref{sec:infrared}.

\subsection{Comparison with previous associations}
\label{sec:compprev}
{ The fraction of sources for which we have been able to find a $\gamma$-ray blazar 
counterpart is about $\sim$15-20\% lower than presented in previous analyses of UGSs \citep{paper4}
and AGUs \citep{paper3}, respectively.
This difference occurs because a more conservative approach has been adopted in the new
parametrization of the $locus$. We not limit blazar candidates to those having 
the {\it scores} higher than 30\% of the entire distribution of $\gamma$-ray emitting blazars \citep{paper6},
rather than 10\% as in the previous analysis.}
{ These choices} made our association method more efficient, so
decreasing the number of \wse\ sources with IR colors similar to those of the $\gamma$-ray blazar population.
In addition, we { now use} a {\it search region} of radius $\theta_{95}$ instead of that at 99.9\% level of confidence,
to be consistent with the associations of the 2FGL and the 2LAC catalogs. 
All the sources listed in this work as $\gamma$-ray blazar candidates were also selected in our previous analysis
based on \wse\ Preliminary data analysis \citep{paper4}.

We note that only three IR \wse\ sources
have the ``contamination and confusion" flag that might indicate a \wse\ spurious detection of an artifact 
in all bands \citep[e.g.,][]{cutri12}.
It occurs for \wse\ J085238.73-575529.4 within the AGUs, \wse\ J084121.63-355505.9 in the UGS sample, 
and WISE J125357.07-583322.3 among the UGS with $\gamma$-ray analysis flags. 
The large majority (i.e., $\sim$90\%) of the \wse\ sources considered do not show any \wse\ analysis flags,
with 10\% clean in at least two IR bands.

Finally, we remark that several $\gamma$-ray pulsars have been identified { since} the release of the 2FGL where they { were} listed as UGSs.
However, we { tested} these UGSs { and} we did not find any { \wse} blazar-like counterpart
associable to them. Thus, in agreement { with other gamma-ray pulsars listed in the} the Public List of LAT-Detected Gamma-Ray Pulsars 
\footnote{\underline{https://confluence.slac.stanford.edu/display/GLAMCOG/Public\\+List+of+LAT-Detected+Gamma-Ray+Pulsars}}.

\section{Correlation with existing databases}
\label{sec:counterparts}
We searched in the following major radio,infrared, optical and X-ray surveys 
as well as in the NASA Extragalactic Database (NED)
\footnote{\underline{http://ned.ipac.caltech.edu/}}
for possible counterparts within 3\arcsec.3 of our $\gamma$-ray blazar candidates, selected with the \wse\
association method, to { see} if additional information { could} confirm their blazar-like nature.
The angular separation of 3\arcsec.3 from the \wse\ position was chosen on the basis of 
the statistical analysis { previously} performed { to assign} a \wse\ counterpart { to each \bzcat\ source}
\citep{paper6} developed following the approach described in Maselli et al. (2012a, 2012b).
{ In particular, we found that for all radii larger than 3\arcsec.3 the increase in the number of IR sources 
positionally associated with \bzcat\ blazars becomes systematically lower 
than the increase in number of random associations. This choice of radius results in
zero multiple matches.}
 
For the radio counterparts we used the NRAO VLA Sky Survey \citep[NVSS;][- N]{condon98}, 
the VLA Faint Images of the Radio Sky at Twenty-Centimeters \citep[FIRST;][- F]{becker95,white97}, 
the Sydney University Molonglo Sky Survey \citep[SUMSS;][- S]{mauch03}
and the Australia Telescope 20 GHz Survey \citep[AT20G;][- A]{murphy10}; 
for the infrared we used the Two Micron All Sky Survey \citep[2MASS;][- M]{skrutskie06}
since each \wse\ source is already associated with the closest 2MASS source 
by the default catalog \citep[see][for more details]{cutri12}.
We also marked sources that are variable { when having}
the {\it variability flag} higher than 5 in at least one band as in the \wse\ all-sky catalog \citep{cutri12}.
Then, we also searched for optical counterparts, with possible spectra available, 
in the Sloan Digital Sky Survey \citep[SDSS; e.g.][- s]{adelman08,paris12}, in the Six-degree-Field Galaxy Redshift Survey 
\citep[6dFGS;][- 6]{jones04,jones09}; 
while for the high energy we looked in the soft X-rays using the ROSAT all-sky survey \citep[RASS;][- X]{voges99}.
A deeper X-ray analysis based on the pointed observations present in the \xmm, \chn, \swf\ and \suz\ archives 
will be performed in a forthcoming paper \citep{paggi13}. We also considered NED for additional information.

We also searched in the USNO-B Catalog \citep{monet03} for the optical counterparts of our 
$\gamma$-ray blazar candidates within 3\arcsec.3; this cross correlation will be useful to prepare 
future follow up observations and the complete list of sources together with their optical magnitudes is reported in Appendix.

\begin{figure}[!ht] 
           \includegraphics[height=9cm,width=8.8cm,angle=0]{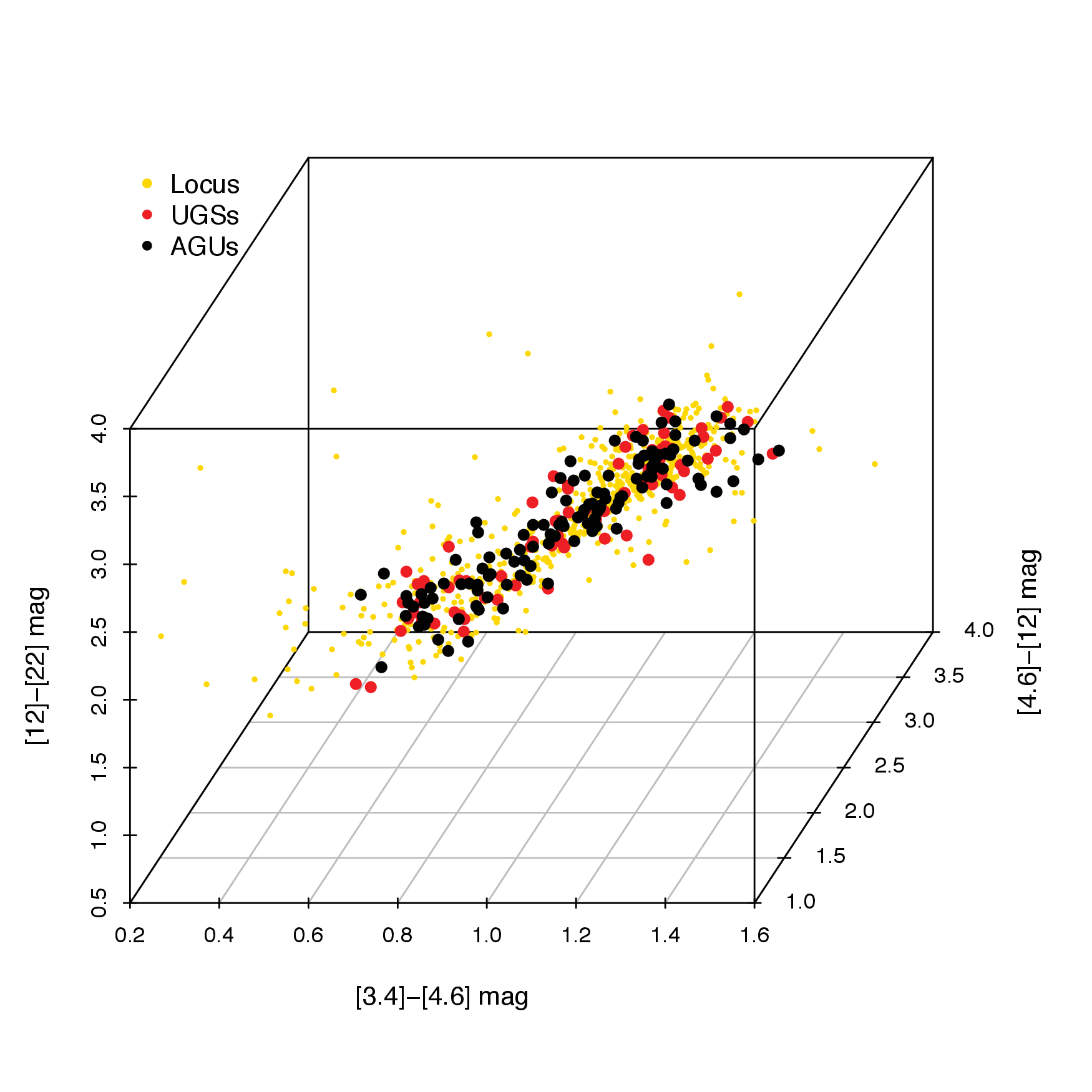}
          \caption{The 3D representation of the $locus$ (known $\gamma$-ray blazars are indicated in yellow)
                         in comparison with the selected $\gamma$-ray blazar candidates: UGSs (red) and AGUs (black).}
          \label{fig:wise}
          \end{figure}

In Table~\ref{tab:sum} we summarize all the multifrequency information for the UGS samples,
without and with the $\gamma$-ray analysis flags, respectively, while all the details are given in 
in Table~\ref{tab:ugs} and Table~\ref{tab:ugf}. In Table~\ref{tab:agu1} and Table~\ref{tab:agu2} we report { our findings} the AGUs.
In each table we report the 2FGL source name, together with that of the \wse\ associated counterpart
and a generic one from the surveys cited above.
We also report the IR \wse\ colors, the type and the class of each candidate derived by our association procedure,
the notes regarding the multifrequency archival analysis, as the optical classification, and, if known, the redshift.
In Table~\ref{tab:agu1} and Table~\ref{tab:agu2}, we also indicate if the selected source
is the same associated by the 2FGL and the 2LAC.  
{ Figure~\ref{fig:wise} shows the 3-dimensional color plot comparing the IR colors of the selected $\gamma$-ray blazar
candidates with the blazar population that constitutes the $locus$.}
\begin{table}
\begin{center}
\caption{Number of counterparts in the radio, infrared, optical and X-rays surveys
for the unidentified gamma-ray sources.}
\begin{tabular}{|lcrr|}
\hline
survey & band & counterparts/total & counterparts/total\\
  &   & UGS (no $\gamma$-flags) & UGSs ($\gamma$-flags)\\
\hline
\noalign{\smallskip}
NVSS           & radio & 19/75 & 4/71\\
FIRST          & radio &  6/75 & 1/71\\
SUMSS          & radio &  7/75 & 0/71\\
2MASS          & infrared & 43/75 &  47/71\\
6dFGS          & optical & 1/75 &  1/71 \\
SDSS           &  optical & 13/75 & 1/71\\
ROSAT           &  X-ray & 3/75 & 1/71 \\
\noalign{\smallskip}
\hline
\end{tabular}\\
\label{tab:sum}
\end{center}
\end{table}

\subsection{Radio counterparts}
\label{sec:radio}
\begin{figure*}[] 
           \includegraphics[height=9.5cm,width=6.5cm,angle=-90]{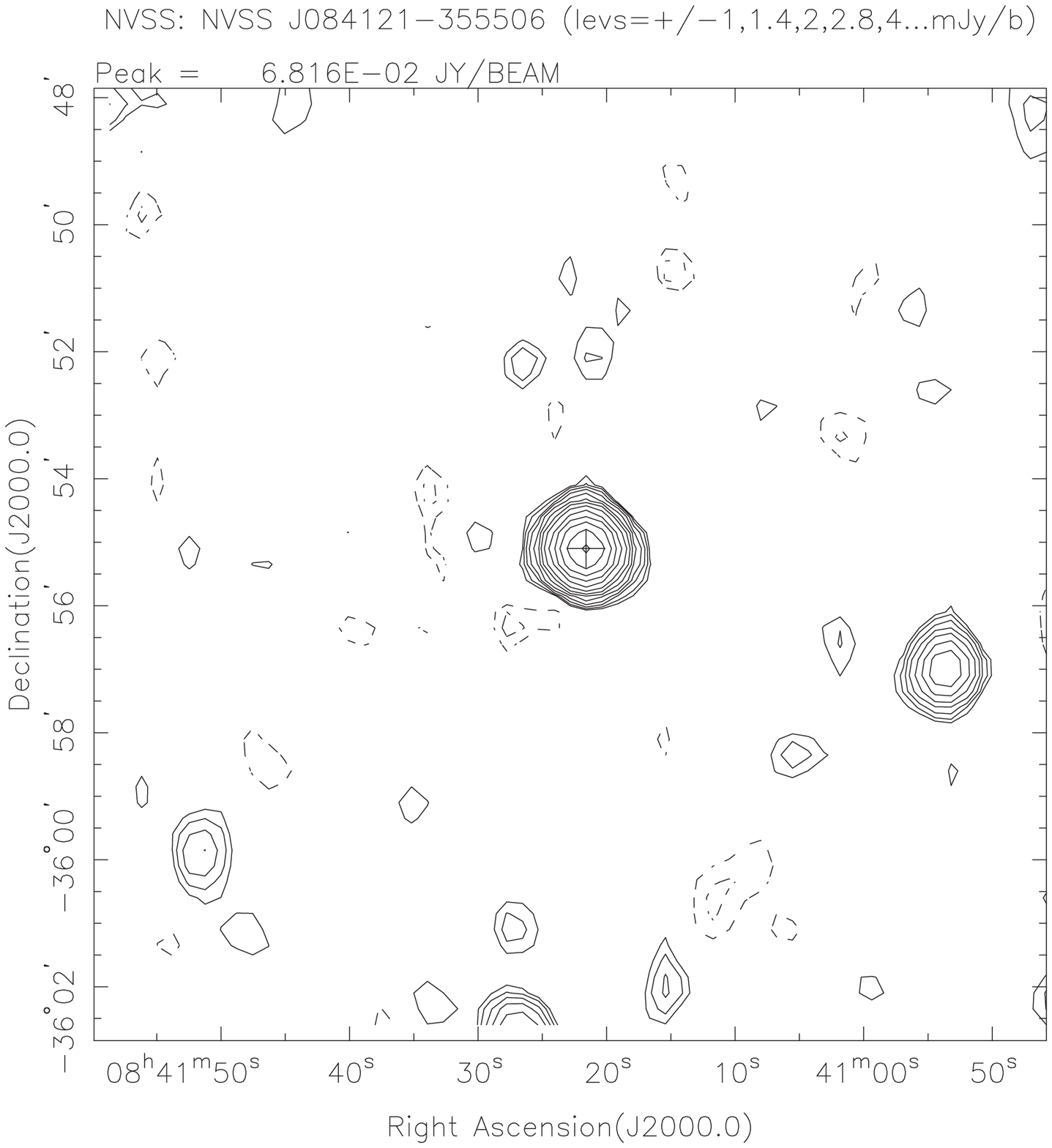}
           \includegraphics[height=9.5cm,width=6.5cm,angle=-90]{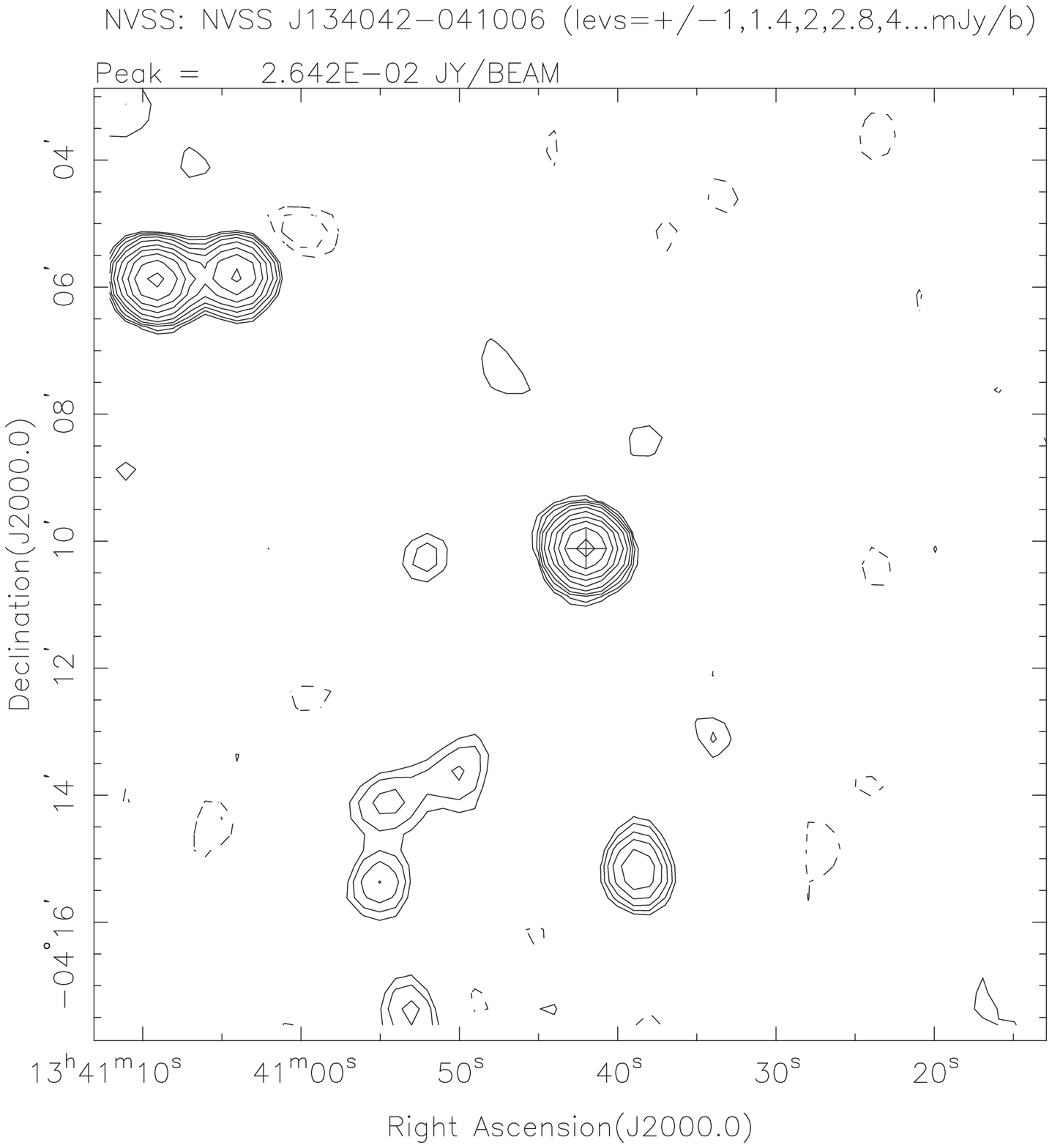}
          \caption{The archival NVSS radio observations (15\arcmin\ radius)
                        of the $\gamma$-ray blazars candidates: \wse\ J084121.63-355505.9 (left) and \wse\ J134042.02-041006.8 (right),
                        associated with the \fer\ sources 2FGLJ0841.3-3556 and 2FGLJ1340.5-0412, respectively. 
                        The black crosses point to the radio counterpart of the $\gamma$-ray blazar candidates selected 
                        according to our association procedure.
                        They are a clear examples of core dominated radio sources similar to blazars in the radio band also at 1.4 GHz.
                        Contour levels are { labeled} together with the NVSS peak flux in Jy/beam.}
          \label{fig:radio}
          \end{figure*}

In the UGS sample of sources without $\gamma$-ray analysis flags, 19 have a counterpart in the NVSS;
7 in the SUMSS and 6 only in the FIRST { (5 in common with the previous 19 in the NVSS)}.
In the list of UGSs with $\gamma$-ray analysis flags, we found only 4 sources { having} a radio counterpart in the NVSS,
one also detected in the FIRST, but none in the SUMSS or in the AT20G catalogs.
In Figure~\ref{fig:radio} we show the archival NVSS radio image of \wse\ J134706.89-295842.3, { the} candidate
low-energy counterpart of 2FGLJ1347.0-2956.

Within the AGU sample, 12 sources out the 38 new associations proposed have unique counterparts
in one of the considered radio survey. Two of them: BZUJ1239+0730 and BZUJ1351-2912,
were also classified as Blazars of uncertain type in the \bzcat\ \citep[e.g.,][]{massaro11}, 
while the remaining one are divided as 6 in the NVSS, 1 in the FIRST,
2 in the SUMSS and 1 in the AT20G.

\subsection{Infrared counterparts}
\label{sec:infrared}
In the UGS sample of sources without $\gamma$-ray analysis flags, there are 43 \wse\ candidates
with counterparts in the 2MASS catalog: 10 out of 75 are variable infrared sources
according to the same criterion previously described. 

The large majority (47 out of 71) of the UGSs, in the sample with $\gamma$-ray analysis flags, 
have counterparts in the 2MASS catalog { and} 15 out of 71 are variable according to the \wse\
all-sky catalog.

{ Of} the 125 \wse\ candidates counterparts of the AGUs, 59 are detected
in 2MASS, as generally expected for blazars \citep[e.g.,][]{chen05}.
In addition, 25 $\gamma$-ray blazar candidates out of 125 have the 
variability flag in the \wse\ catalog with a value higher than 5 in at least one band,
suggesting that their IR emission { is not likely} arising from dust.

\subsection{Optical counterparts}
\label{sec:optical}
In the sample of UGSs without $\gamma$-ray analysis flags, 13 sources have been found with a counterpart in the SDSS,
4 with spectroscopic information (Table~\ref{tab:ugs}).
Among these 4 sources, two are broad line quasars, promising to be blazar-like sources of BZQ type.
One is a Seyfert galaxy: SDSS J015910.05+010514.5 is a contaminant of our association procedure
(although our method suggests a better candidate, for 2FGLJ0158.4+0107).
The remaining one, NVSS J161543+471126 shows the optical spectrum similar to that of an
X-ray Bright, Optically Normal Galaxy \citep[XBONG][]{comastri02}.  
The source SDSS J015836.23+010632.0, another candidate counterpart of 2FGLJ0158.4+0107 
is described as a quasar at redshift 0.723 in Schneider et al. (2007) and Hu et al. (2008).
In addition to these 4 sources, spectroscopic information is also available for WISE J230010.16-360159.9
a possible low-energy counterpart of 2FGLJ2300.0-3553, classified as quasar according to Jones et al. (2009).
A quasar-like spectrum is then available for SDSS J161434.67+470420.0 candidate counterpart of 2FGLJ1614.8+4703.

\begin{figure}[!hb] 
           \includegraphics[height=9.6cm,width=5.6cm,angle=-90]{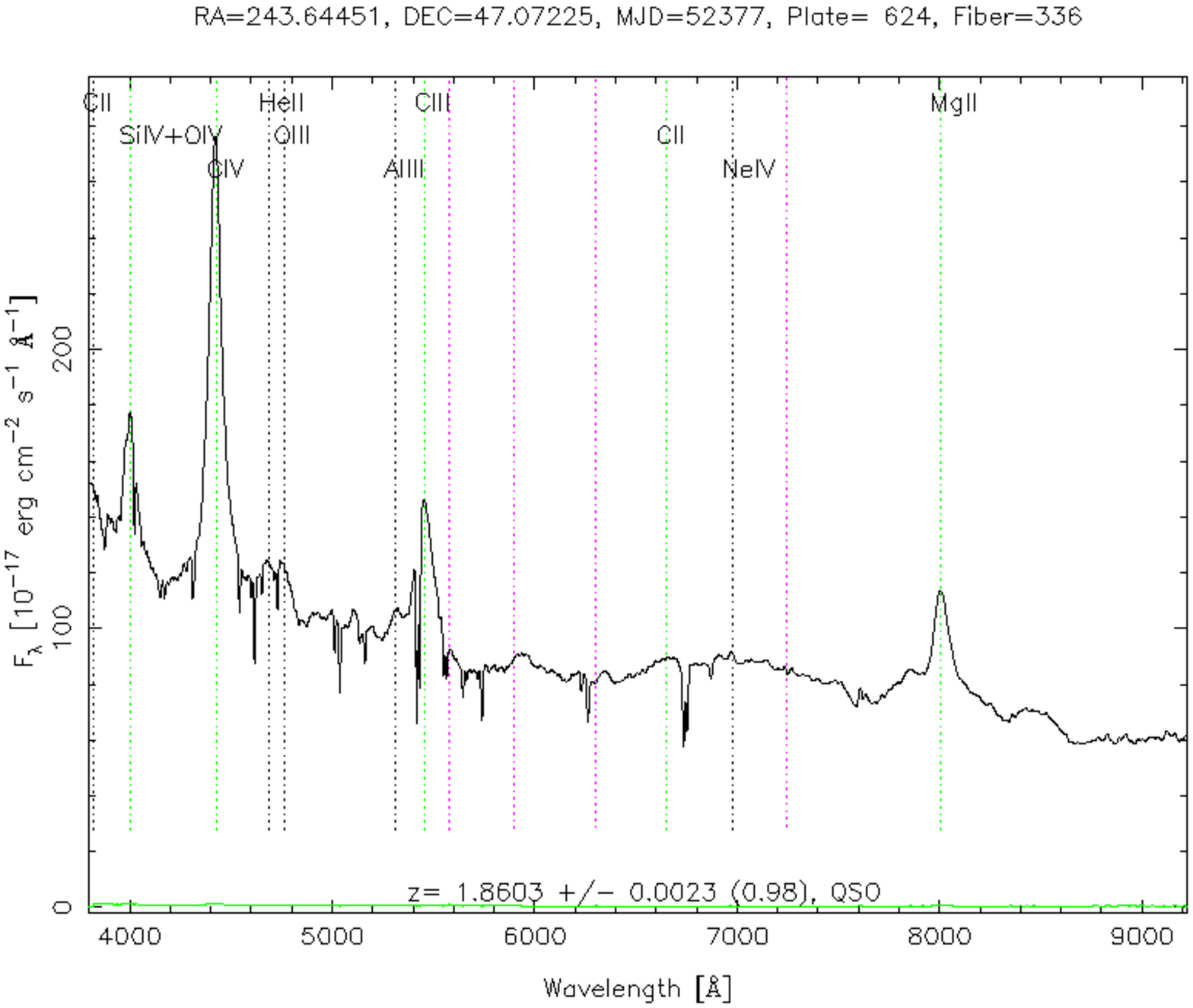}
          \caption{The archival SDSS spectroscopic observation 
                        of the $\gamma$-ray blazars candidate \wse\ J161434.67+470420.1
                        associated with the \fer\ source 2FGLJ1614.8+4703. 
                        This optical spectrum { indicates} toward a BZQ classification of \wse\ J161434.67+470420.1.}
          \label{fig:optical2}
          \end{figure}

The search for the optical counterparts for UGSs with $\gamma$-ray analysis flags was less successful.
Only one source has an optical, counterpart: WISE J131552.98-073301.9, associated with 2FGLJ1315.6-0730.
This source has a counterpart in both the NVSS and in the FIRST radio survey.
According to Bauer et al. (2009), this source is also variable in the optical and it was therefore selected as a blazar candidate.
In Figure~\ref{fig:optical2} we show the archival SDSS spectrum of the \wse\ J161434.67+470420.1 candidate
{ as the} low energy counterpart of 2FGLJ1614.8+4703.

We found only 1 $\gamma$-ray blazar candidate in the AGU sample with a counterpart in the SDSS,
while 4 of them have a 6dFGS source lying  3\arcsec.3 from their \wse\ position.
In the case of WISE J033200.72-111456.1 associated with 2FGLJ0332.5-1118, we also found that its 6dFGS
optical spectrum appear to be featureless suggesting { a} BL Lac classification \citep{jones09}.
The same information has been found for WISE J001920.58-815251.3 associated with 2FGLJ0018.8-8154, 
for which the noisy, featureless 6dFGS optical spectrum points to a BL Lac classification \citep{jones09}. 
2FGLJ0823.0+4041 and 2FGLJ0858.1-1952 appear to be associated, both by the 2FGL catalog and our method to broad line quasars.
WISE J085805.36-195036.8 associated with 2FGLJ0858.1-1952 is also classified as a quasar at redshift 0.6597
by White et al. (1988). 
The archival 6dFGS spectrum of \wse\ J001920.58-815251.3 { the} candidate
low energy counterpart of 2FGLJ0018.8-8154 { is available on NED}; the absence of features allows { us} to classify the source as a BZB.

\subsection{X-ray counterparts}
\label{sec:xray}
In the UGS sample without $\gamma$-ray analysis flags, only 3 objects have X-ray counterparts in the ROSAT all-sky catalog:
the Seyfert 1 galaxy SDSS J015910.05+010514.5, the quasar SDSS J161434.67+470420.0 
(both described in Section~\ref{sec:optical}) and WISE J164619.95+435631.0 associated with 2FGLJ1647.0+4351.
In addition, SDSS J161434.67+470420.0 is also detected in the \chn\ source catalog: CXO J161434.7+470419 { as occurs}
NVSS J161543+471126, alias CXO J161541.2+471111 \citep{evans10}.

In the UGS list of sources with $\gamma$-ray analysis flags, { there is} only a single object detected in the ROSAT
all-sky survey, namely WISE  J043947.48+260140.5 uniquely associated with the \fer\ source 2FGLJ0440.5+2554c 
and with a X-ray counterpart also in the \chn\ source catalog CXO J043947.5+260140 \citep{evans10}.
In addition, WISE J060659.94-061641.5 { the} unique counterpart of 2FGLJ0607.5-0618c,
has the \chn\ counterpart CXO J060700.1-061641 \citep{evans10}.

Finally, in the AGU sample of 38 new $\gamma$-ray blazar candidates we found only 1 source in the ROSAT catalog, namely 
WISE J181037.99+533501.5, associated with the X-ray object 1RXS J181038.5+533458 
and having a radio counterpart in the NVSS.
According to NED, WISE J182352.33+431452.5, associated with 2FGLJ1823.8+4312,
is also detected in the X-rays by \chn: CXO J182352.2+431452 \citep{paper7}.

\begin{table*}
\caption{Unidentified Gamma-ray Sources without $\gamma$-ray analysis flags.}
\tiny
\begin{tabular}{|lllccccclc|}
\hline
  2FGL  &  \wse\  &  other & [3.4]-[4.6] & [4.6]-[12] & [12]-[22] & type & class & notes & z \\
  name  &  name   &  name  &     mag     &    mag     &   mag     &      &       &       &   \\
\hline
\noalign{\smallskip}
J0039.1+4331  &  J003858.27+432947.0  &                        &  0.98(0.04)  &  2.36(0.05)  &  2.30(0.15)  &  BZB  &  C  & M         & ?    \\
J0116.6-6153  &  J011619.59-615343.5  &  SUMSS J011619-615343  &  0.85(0.04)  &  2.34(0.06)  &  1.85(0.25)  &  BZB  &  C  & S         & ?    \\
J0133.4-4408  &  J013321.36-441319.4  &                        &  1.12(0.05)  &  3.20(0.06)  &  2.62(0.15)  &  BZQ  &  C  &           & ?    \\ 
              &  J013306.35-441421.3  &  SUMSS J013306-441422  &  0.83(0.03)  &  2.25(0.05)  &  1.92(0.22)  &  BZB  &  C  & S         & ?    \\
J0143.6-5844  &  J014347.39-584551.3  &  SUMSS J014347-584550  &  0.68(0.03)  &  1.93(0.06)  &  1.89(0.29)  &  BZB  &  C  & S,M       & ?    \\
J0158.4+0107  &  J015836.25+010632.1  &SDSS J015836.23+010632.0&  1.05(0.03)  &  2.50(0.05)  &  2.33(0.11)  &  UND  &  C  & M,s,QSO   & 0.723?\\
              &  J015757.45+011547.8  &                        &  1.06(0.04)  &  2.52(0.06)  &  2.18(0.18)  &  UND  &  C  &           & ?    \\
              &  J015910.05+010514.7  &SDSS J015910.05+010514.5&  0.91(0.03)  &  2.47(0.03)  &  2.48(0.05)  &  UND  &  C  & M,s,X,Sy1 & 0.217\\
J0158.6+8558  &  J015550.16+854745.1  &                        &  1.12(0.04)  &  2.50(0.06)  &  2.23(0.18)  &  UND  &  C  &           & ?    \\
              &  J014935.30+860115.3  &                        &  0.68(0.03)  &  2.32(0.05)  &  2.00(0.17)  &  BZB  &  C  & M         & ?    \\
J0227.7+2249  &  J022744.35+224834.3  &  NVSS J022744+224834   &  0.93(0.03)  &  2.58(0.03)  &  2.09(0.08)  &  BZB  &  B  & N,M,v     & ?    \\
J0316.1-6434  &  J031614.31-643731.4  &  SUMSS J031614-643732  &  0.74(0.03)  &  2.09(0.06)  &  1.84(0.26)  &  BZB  &  C  & S,M       & ?    \\
J0332.1+6309  &  J033153.90+630814.1  &  NVSS J033153+630814   &  0.77(0.03)  &  2.38(0.04)  &  1.96(0.11)  &  BZB  &  B  & N,M       & ?    \\
J0409.8-0357  &  J040946.57-040003.4  &  NVSS J040946-040003   &  0.88(0.03)  &  2.36(0.04)  &  1.94(0.12)  &  BZB  &  B  & N,M       & ?    \\
J0414.9-0855  &  J041457.01-085652.0  &  APMUKS 041232.66-090420.3    &  1.00(0.04)  &  2.63(0.08)  &  2.31(0.23)  &  UND  &  C  &           & ?    \\
J0416.0-4355  &  J041605.81-435514.6  &  SUMSS J041605-435516  &  1.11(0.03)  &  2.89(0.04)  &  2.44(0.08)  &  BZQ  &  B  & S,M       & ?    \\
J0431.5+3622  &  J043103.34+362158.7  &                        &  1.14(0.06)  &  2.72(0.11)  &  2.44(0.26)  &  BZQ  &  C  &           & ?    \\ 
J0555.9-4348  &  J055531.59-435030.7  &                        &  1.36(0.05)  &  3.14(0.06)  &  2.39(0.18)  &  BZQ  &  C  &           & ?    \\ 
              &  J055618.74-435146.1  &  SUMSS J055618-435146  &  0.90(0.03)  &  2.49(0.04)  &  2.14(0.13)  &  BZB  &  B  & S,M       & ?    \\
J0602.7-4011  &  J060237.10-401453.2  &                        &  0.96(0.03)  &  2.47(0.03)  &  2.35(0.06)  &  UND  &  B  & M         & ?    \\
J0644.6+6034  &  J064459.38+603131.7  &                        &  0.97(0.04)  &  2.59(0.06)  &  2.50(0.15)  &  UND  &  C  & M         & ?    \\
J0713.5-0952  &  J071223.28-094536.3  &                        &  1.16(0.04)  &  3.12(0.05)  &  2.27(0.08)  &  BZQ  &  C  & M         & ?    \\
J0723.9+2901  &  J072354.83+285929.9  &  NVSS J072354+285930   &  1.14(0.05)  &  2.89(0.05)  &  2.40(0.11)  &  BZQ  &  C  & N,F       & ?    \\
J0744.1-2523  &  J074401.10-252205.9  &                        &  1.09(0.04)  &  2.57(0.03)  &  2.69(0.05)  &  BZQ  &  C  &           & ?    \\
              &  J074402.19-252146.0  &                        &  1.00(0.03)  &  2.00(0.03)  &  2.16(0.03)  &  BZB  &  B  & M         & ?    \\
J0746.0-0222  &  J074627.03-022549.3  &  NVSS J074627-022549   &  0.67(0.04)  &  2.09(0.07)  &  1.99(0.30)  &  BZB  &  C  & N,M       & ?    \\
J0756.3-6433  &  J075624.60-643030.6  &                        &  0.87(0.03)  &  2.21(0.06)  &  2.11(0.18)  &  BZB  &  C  & M,v       & ?    \\
J0807.0-6511  &  J080729.66-650910.3  &                        &  1.12(0.05)  &  3.08(0.07)  &  2.46(0.17)  &  BZQ  &  C  &           & ?    \\ 
J0838.8-2828  &  J083842.77-282830.9  &                        &  1.03(0.05)  &  2.73(0.09)  &  2.24(0.27)  &  UND  &  C  &           & ?    \\ 
J0841.3-3556  &  J084121.63-355505.9  &  NVSS J084121-355506   &  0.79(0.03)  &  2.23(0.03)  &  1.77(0.11)  &  BZB  &  B  & N,M       & ?    \\
J0844.9+6214  &  J084406.81+621458.6  &SDSS J084406.83+621458.5&  0.68(0.03)  &  2.07(0.05)  &  2.23(0.14)  &  BZB  &  C  & M,s       & ?    \\
J0855.4-4625  &  J085548.43-462244.3  &                        &  1.19(0.03)  &  2.32(0.03)  &  2.16(0.04)  &  UND  &  C  &           & ?    \\
J0858.3-4333  &  J085839.22-432642.7  &                        &  0.74(0.03)  &  1.82(0.03)  &  2.29(0.04)  &  BZB  &  C  & M         & ?    \\
J0900.9+6736  &  J090121.65+673955.8  &                        &  0.94(0.05)  &  2.84(0.08)  &  2.16(0.26)  &  UND  &  C  & M         & ?    \\ 
J0955.0-3949  &  J095458.30-394655.0  &                        &  0.77(0.04)  &  2.28(0.05)  &  2.03(0.18)  &  BZB  &  C  & M         & ?    \\
J1013.6+3434  &  J101256.54+343648.8  &SDSS J101256.54+343648.7&  0.89(0.04)  &  2.62(0.07)  &  2.09(0.26)  &  BZB  &  C  & M,s       & ?    \\
J1016.1+5600  &  J101544.44+555100.7  &  NVSS J101544+555100   &  1.05(0.06)  &  3.08(0.09)  &  2.57(0.25)  &  BZQ  &  C  & N,F,s     & ?    \\ 
J1029.5-2022  &  J102946.66-201812.6  &                        &  1.26(0.06)  &  2.88(0.10)  &  2.59(0.26)  &  BZQ  &  C  & M         & ?    \\ 
J1032.9-8401  &  J103015.35-840308.7  &  SUMSS J103014-840307  &  0.96(0.04)  &  2.59(0.05)  &  2.07(0.15)  &  BZB  &  C  & S,v       & ?    \\
J1038.2-2423  &  J103754.92-242544.5  &                        &  1.22(0.04)  &  3.26(0.04)  &  2.58(0.10)  &  BZQ  &  C  & M         & ?    \\ 
J1207.3-5055  &  J120750.50-510314.9  &                        &  1.05(0.05)  &  2.97(0.08)  &  2.55(0.19)  &  BZQ  &  C  &           & ?    \\ 
              &  J120746.43-505948.6  &                        &  1.14(0.05)  &  2.87(0.07)  &  2.53(0.15)  &  BZQ  &  C  &           & ?    \\ 
J1254.2-2203  &  J125422.47-220413.6  &  NVSS J125422-220413   &  0.66(0.04)  &  2.31(0.08)  &  1.77(0.36)  &  BZB  &  C  & N,M,v     & ?    \\
J1259.8-3749  &  J125949.80-374858.1  &  NVSS J125949-374856   &  0.70(0.04)  &  2.10(0.08)  &  1.99(0.33)  &  BZB  &  C  & N,S,M,v   & ?    \\
J1340.5-0412  &  J134042.02-041006.8  &  NVSS J134042-041006   &  0.71(0.04)  &  2.11(0.08)  &  1.85(0.41)  &  BZB  &  C  & N,v       & ?    \\
J1346.0-2605  &  J134621.08-255642.3  &                        &  1.14(0.06)  &  2.94(0.10)  &  2.58(0.26)  &  BZQ  &  C  &           & ?    \\ 
J1347.0-2956  &  J134706.89-295842.3  &  NVSS J134706-295840   &  0.78(0.03)  &  2.10(0.06)  &  1.91(0.23)  &  BZB  &  C  & N,S,M,v   & ?    \\
J1404.0-5244  &  J140313.11-524839.5  &                        &  1.22(0.05)  &  2.97(0.07)  &  2.63(0.15)  &  BZQ  &  C  &           & ?    \\ 
J1517.2+3645  &  J151649.26+365022.9  &  NVSS J151649+365023   &  0.95(0.03)  &  2.63(0.04)  &  2.07(0.12)  &  BZB  &  C  & N,F,s,v   & ?    \\
J1612.0+1403  &  J161118.10+140328.9  &SDSS J161118.10+140328.7&  1.06(0.06)  &  3.15(0.09)  &  2.55(0.22)  &  BZQ  &  C  & s         & ?    \\ 
J1614.8+4703  &  J161450.96+465953.7  &SDSS J161450.91+465953.6&  1.20(0.04)  &  2.78(0.06)  &  2.54(0.16)  &  BZQ  &  C  & s         & ?    \\ 
              &  J161513.04+471355.2  &SDSS J161513.04+471355.4&  1.17(0.05)  &  2.86(0.10)  &  2.61(0.24)  &  BZQ  &  C  & s         & ?    \\ 
              &  J161434.67+470420.1  &SDSS J161434.67+470420.0&  1.08(0.03)  &  3.12(0.03)  &  2.19(0.04)  &  BZQ  &  C  & F,M,s,X,x,QSO&1.86\\
              &  J161541.22+471111.8  &  NVSS J161543+471126   &  0.74(0.03)  &  2.28(0.04)  &  2.28(0.09)  &  BZB  &  C  & N,F,M,s,x,XB?&0.199\\
J1622.8-0314  &  J162225.35-031439.6  &                        &  1.10(0.06)  &  3.19(0.09)  &  2.67(0.18)  &  BZQ  &  C  &           & ?    \\ 
J1627.8+3219  &  J162800.40+322414.0  &SDSS J162800.39+322413.9&  1.13(0.05)  &  2.83(0.10)  &  2.62(0.24)  &  BZQ  &  C  & s         & ?    \\ 
J1647.0+4351  &  J164619.95+435631.0  &  NVSS J164619+435631   &  0.77(0.04)  &  2.09(0.09)  &  2.11(0.39)  &  BZB  &  C  & N,F,s,X   & ?    \\ 
J1730.6-0353  &  J173052.86-035247.1  &                        &  1.16(0.04)  &  2.92(0.04)  &  2.28(0.07)  &  BZQ  &  B  & M         & ?    \\
J1742.5-3323  &  J174201.11-332607.9  &                        &  0.64(0.05)  &  1.73(0.03)  &  1.60(0.03)  &  BZB  &  B  & M         & ?    \\
J1745.6+0203  &  J174507.82+015442.5  &  NVSS J174507+015445   &  1.26(0.03)  &  3.40(0.03)  &  2.45(0.03)  &  BZQ  &  A  & N,M       & ?    \\
              &  J174526.95+020532.7  &                        &  0.94(0.03)  &  2.57(0.03)  &  2.27(0.07)  &  UND  &  B  & M         & ?    \\
J1759.2-3853  &  J175903.29-384739.5  &                        &  0.58(0.04)  &  1.93(0.03)  &  1.50(0.02)  &  BZB  &  A  & M         & ?    \\
J1842.3+2740  &  J184201.25+274239.2  &                        &  1.22(0.06)  &  3.03(0.08)  &  2.43(0.22)  &  BZQ  &  C  &           & ?    \\ 
J1904.8-0705  &  J190444.57-070740.1  &                        &  0.91(0.05)  &  2.79(0.09)  &  2.45(0.19)  &  UND  &  C  & M         & ?    \\
J1924.9-1036  &  J192501.63-104316.3  &                        &  1.24(0.05)  &  3.25(0.05)  &  2.66(0.09)  &  BZQ  &  C  & M         & ?    \\
J2004.6+7004  &  J200503.41+700236.3  &                        &  1.17(0.05)  &  2.95(0.07)  &  2.21(0.24)  &  BZQ  &  C  &           & ?    \\ 
              &  J200506.02+700439.3  &  NVSS J200506+700440   &  0.70(0.03)  &  2.11(0.05)  &  2.11(0.18)  &  BZB  &  C  & N,v       & ?    \\
J2021.5+0632  &  J202154.66+062908.7  &                        &  0.96(0.04)  &  2.67(0.06)  &  2.47(0.10)  &  UND  &  C  & M         & ?    \\
              &  J202155.45+062913.7  &  NVSS J202155+062914   &  0.80(0.03)  &  2.09(0.05)  &  1.78(0.17)  &  BZB  &  C  & N,M       & ?    \\
J2114.1+5440  &  J211508.92+544815.7  &                        &  1.11(0.03)  &  2.50(0.03)  &  2.52(0.04)  &  BZQ  &  B  & M         & ?    \\
J2133.9+6645  &  J213349.21+664704.3  &  NVSS J213349+664706   &  0.67(0.04)  &  2.12(0.06)  &  1.86(0.22)  &  BZB  &  C  & N,v       & ?    \\
J2134.6-2130  &  J213430.18-213032.6  &  NVSS J213430-213032   &  0.77(0.04)  &  2.26(0.08)  &  1.78(0.42)  &  BZB  &  C  & N,M       & ?    \\
J2300.0-3553  &  J230010.16-360159.9  &  6dF J2300101-360200   &  1.17(0.06)  &  3.36(0.08)  &  2.43(0.22)  &  BZQ  &  C  & M,6,QSO   & 2.357\\ 
J2319.3-3830  &  J232000.11-383511.4  &  MRSS 347-103293       &  1.12(0.05)  &  3.06(0.07)  &  2.60(0.16)  &  BZQ  &  C  &           & ?    \\ 
J2358.4-1811  &  J235828.61-181526.6  &                        &  1.09(0.04)  &  2.60(0.06)  &  2.46(0.15)  &  UND  &  C  & M         & ?    \\
\noalign{\smallskip}
\hline
\end{tabular}\\
Col. (1) 2FGL name. \\
Col. (2) \wse\ name. \\
Col. (3) Other name if present in literature and in the following order: \bzcat, NVSS, SDSS, AT20G, NED. \\
Cols. (4,5,6) Infrared colors from the \wse\ all sky catalog corrected for Galactic extinciton. Values in parentheses are 1$\sigma$ uncertainties. \\
Col. (7) Type of candidate according to our method: BZB - BZQ - UND (undetermined). \\
Col. (8) Class of candidate according to our method. \\
Col. (9) Notes: N = NVSS, F = FIRST, S = SUMSS, A=AT20G, M = 2MASS, s = SDSS dr9, 6 = 6dFGS, x = \xmm\ or \chn, X = ROSAT;  QSO  = quasar, Sy = Seyfert, LNR = LINER, BL = BL Lac, XB = X-ray Bright Optically Inactive Galaxies; 
v = variability in \wse\ (var\_flag $>$ 5 in at least one band). \\
Col. (10) Redshift: (?) = unknown, (number?) = uncertain. 
\label{tab:ugs}
\end{table*}

\begin{table*}
\caption{Unidentified Gamma-ray Sources with $\gamma$-ray analysis flags.}
\tiny
\begin{tabular}{|lllccccclc|}
\hline
  2FGL  &  \wse\  &  other & [3.4]-[4.6] & [4.6]-[12] & [12]-[22] & type & class & notes & z \\
  name  &  name   &  name  &     mag     &    mag     &   mag     &      &       &       &   \\
\hline
\noalign{\smallskip}
J0233.9+6238c &  J023418.09+624207.8  &                           &  0.80(0.03) &  2.02(0.04) &  2.31(0.06) &  BZB  &  C  &  M    &  ?\\
              &  J023238.07+623651.9  &                           &  1.25(0.05) &  3.05(0.06) &  2.46(0.11) &  BZQ  &  C  &       &  ?\\ 
J0341.8+3148c &  J034204.35+314711.4  &                           &  0.48(0.03) &  1.70(0.03) &  1.78(0.08) &  BZB  &  B  &  M    &  ?\\
              &  J034141.11+314804.5  &                           &  0.93(0.03) &  2.47(0.05) &  2.20(0.13) &  UND  &  C  &  M,v  &  ?\\
              &  J034158.52+314855.7  &                           &  0.64(0.04) &  1.37(0.03) &  1.50(0.02) &  BZB  &  C  &  M    &  ?\\
J0423.4+5612  &  J042430.44+561525.8  &                           &  1.19(0.06) &  2.98(0.13) &  2.47(0.26) &  BZQ  &  C  &       &  ?\\ 
J0440.5+2554c &  J043947.48+260140.5  &  CXO J043947.5+260140     &  0.47(0.03) &  1.47(0.03) &  1.90(0.04) &  BZB  &  A  &  M,X,x&  ?\\
J0543.2-0120c &  J054324.78-011545.6  &                           &  0.69(0.03) &  1.64(0.04) &  1.86(0.11) &  BZB  &  B  &  M    &  ?\\
J0547.5-0141c &  J054758.25-013616.6  &                           &  1.06(0.04) &  2.60(0.06) &  2.20(0.19) &  UND  &  C  &       &  ?\\
J0607.5-0618c &  J060659.94-061641.5  &  CXO J060700.1-061641     &  1.17(0.04) &  2.45(0.07) &  2.33(0.18) &  ?    &  C  & x     &  ?\\
J0616.6+2425  &  J061549.51+241654.0  &                           &  1.12(0.06) &  2.94(0.09) &  2.44(0.23) &  BZQ  &  C  &       &  ?\\ 
              &  J061641.04+241138.4  &                           &  0.87(0.03) &  1.76(0.07) &  1.88(0.23) &  BZB  &  C  &  M    &  ?\\
J0620.8-2556  &  J062108.68-255757.9  &  NVSS J062108-255757      &  0.95(0.05) &  2.75(0.08) &  2.26(0.26) &  ?    &  C  &  N    &  ?\\ 
J0631.7+0428  &  J063104.12+042012.6  &                           &  1.08(0.03) &  2.44(0.03) &  2.56(0.03) &  BZQ  &  A  &  M    &  ?\\
J0634.3+0356c &  J063459.32+040808.6  &                           &  1.13(0.03) &  2.11(0.03) &  2.73(0.04) &  BZQ  &  C  &  M,v  &  ?\\
              &  J063519.84+035047.1  &                           &  1.36(0.04) &  2.73(0.05) &  2.36(0.08) &  BZQ  &  C  &  M    &  ?\\
J0637.0+0416c &  J063705.96+042537.2  &                           &  0.82(0.03) &  2.17(0.04) &  2.11(0.06) &  BZB  &  B  &  M    &  ?\\
              &  J063701.93+042037.2  &                           &  0.86(0.03) &  2.17(0.03) &  2.17(0.03) &  BZB  &  A  &  M,v  &  ?\\
              &  J063703.09+042146.1  &                           &  0.60(0.03) &  1.99(0.07) &  2.06(0.12) &  BZB  &  C  &  M,v  &  ?\\
              &  J063647.19+042058.7  &                           &  0.52(0.03) &  1.79(0.05) &  1.93(0.18) &  BZB  &  C  &  M    &  ?\\
J0708.5-1020c &  J070807.98-102743.9  &                           &  0.55(0.03) &  1.55(0.05) &  1.97(0.11) &  BZB  &  C  &  M    &  ?\\
              &  J070809.69-102805.8  &                           &  1.35(0.03) &  2.91(0.03) &  2.50(0.04) &  BZQ  &  B  &       &  ?\\
              &  J070806.04-102736.2  &                           &  0.62(0.03) &  1.87(0.03) &  2.07(0.05) &  BZB  &  B  &  M    &  ?\\
              &  J070813.96-102840.2  &                           &  0.89(0.03) &  1.73(0.05) &  2.21(0.09) &  BZB  &  C  &       &  ?\\
              &  J070826.05-103001.2  &                           &  1.18(0.04) &  2.88(0.04) &  2.48(0.05) &  BZQ  &  B  &  M,v  &  ?\\
              &  J070816.06-102832.0  &                           &  0.72(0.03) &  1.51(0.06) &  1.95(0.18) &  BZB  &  C  &  M    &  ?\\
J0742.7-3113  &  J074303.55-312057.3  &                           &  1.04(0.05) &  2.57(0.10) &  2.22(0.34) &  UND  &  C  &       &  ?\\ 
J0748.5-2204  &  J074835.46-215740.0  &                           &  1.32(0.05) &  2.80(0.07) &  2.68(0.14) &  BZQ  &  C  &       &  ?\\ 
J0900.5-4441c &  J090039.83-443510.0  &  2MASX J09003978-4435094  &  1.08(0.03) &  1.48(0.03) &  2.32(0.02) &  BZB  &  C  &  M,v  &  ?\\
J0914.1-4756  &  J091319.84-475730.8  &                           &  1.08(0.03) &  1.71(0.03) &  2.20(0.03) &  BZB  &  A  &       &  ?\\
J0922.2-5214c &  J092154.24-521236.1  &                           &  1.19(0.03) &  2.43(0.03) &  2.52(0.02) &  BZQ  &  A  &  M    &  ?\\
J1027.4-5730c &  J102703.87-572830.7  &                           &  1.14(0.04) &  2.81(0.05) &  2.59(0.06) &  BZQ  &  B  &  M,v  &  ?\\
J1059.9-2051  &  J110025.72-205333.4  &  2MASX J11002568-2053333  &  0.89(0.03) &  2.46(0.05) &  1.99(0.14) &  BZB  &  B  &  M    &  ?\\
J1208.6-2257  &  J120816.33-224921.9  &                           &  0.98(0.04) &  2.25(0.05) &  2.00(0.18) &  BZB  &  C  &  M    &  ?\\
J1248.6-5510  &  J124946.07-550758.6  &                           &  0.95(0.04) &  2.71(0.05) &  2.52(0.12) &  UND  &  C  &       &  ?\\
J1255.8-5828  &  J125459.44-582009.5  &                           &  1.20(0.03) &  2.61(0.04) &  2.25(0.05) &  UND  &  C  &  M    &  ?\\
              &  J125357.07-583322.3  &                           &  0.62(0.04) &  2.04(0.03) &  1.56(0.02) &  BZB  &  A  &  M,v  &  ?\\
J1315.6-0730  &  J131543.62-073659.0  &                           &  0.93(0.04) &  2.39(0.06) &  2.15(0.21) &  BZB  &  C  &  M    &  ?\\
              &  J131552.98-073301.9  &  NVSS J131552-073301      &  0.87(0.03) &  2.27(0.04) &  2.04(0.08) &  BZB  &  B  &  N,F,M,6,v,BL?  &  ?\\
J1324.4-5411  &  J132415.49-541104.4  &                           &  1.33(0.05) &  2.88(0.07) &  2.46(0.18) &  BZQ  &  C  &       &  ?\\ 
J1345.8-3356  &  J134543.05-335643.3  &  NVSS J134543-335643      &  0.82(0.04) &  2.31(0.06) &  2.09(0.20) &  BZB  &  C  &  N,M  &  ?\\
J1407.4-2948  &  J140818.86-294203.2  &                           &  1.10(0.06) &  2.82(0.12) &  2.70(0.28) &  BZQ  &  C  &       &  ?\\ 
J1512.5-6247c &  J151156.72-625231.1  &                           &  1.20(0.10) &  2.68(0.09) &  2.68(0.18) &  BZQ  &  C  &       &  ?\\ 
J1624.2-2124  &  J162343.89-210707.0  &                           &  1.34(0.05) &  2.86(0.08) &  2.74(0.16) &  BZQ  &  C  &       &  ?\\ 
J1632.6-2328c &  J163306.53-233207.3  &                           &  1.07(0.03) &  2.71(0.04) &  2.43(0.09) &  UND  &  C  &  M    &  ?\\
J1639.8-4921c &  J163907.38-492605.6  &                           &  0.64(0.05) &  1.66(0.03) &  1.64(0.03) &  BZB  &  B  &  M,v  &  ?\\
J1716.6-0526c &  J171717.66-052520.1  &                           &  1.25(0.04) &  2.95(0.04) &  2.37(0.06) &  BZQ  &  B  &  M    &  ?\\
J1747.2-3507  &  J174741.23-350334.3  &                           &  0.58(0.05) &  1.80(0.03) &  1.41(0.02) &  BZB  &  B  &  M    &  ?\\
J1749.7-3134c &  J174943.85-314054.7  &                           &  0.92(0.03) &  1.15(0.03) &  1.74(0.02) &  BZB  &  C  &  M,v  &  ?\\
              &  J174949.84-313045.6  &                           &  1.10(0.03) &  1.95(0.03) &  2.30(0.02) &  ?    &  A  &  M    &  ?\\
J1754.1-2930  &  J175414.40-293326.6  &                           &  0.68(0.06) &  1.81(0.04) &  1.48(0.05) &  BZB  &  C  &  M    &  ?\\
J1829.8-0204c &  J182927.72-020531.5  &                           &  0.55(0.03) &  1.52(0.03) &  1.94(0.04) &  BZB  &  B  &  M    &  ?\\
              &  J182958.23-015805.1  &                           &  0.71(0.03) &  1.24(0.03) &  1.77(0.04) &  BZB  &  A  &  M    &  ?\\
              &  J183009.32-020723.2  &                           &  0.75(0.03) &  1.48(0.04) &  1.84(0.06) &  BZB  &  B  &  M,v  &  ?\\
J1835.4+1036  &  J183551.92+103056.8  &                           &  0.94(0.03) &  2.82(0.04) &  2.56(0.06) &  UND  &  C  &  M    &  ?\\
J1835.4+1349  &  J183522.00+135733.9  &                           &  1.12(0.04) &  2.61(0.05) &  2.20(0.14) &  UND  &  C  &  M    &  ?\\
              &  J183535.34+134848.8  &                           &  0.65(0.04) &  2.14(0.05) &  2.10(0.15) &  BZB  &  C  &  M    &  ?\\
J1837.9+3821  &  J183828.80+382704.3  &                           &  1.04(0.05) &  2.64(0.09) &  2.52(0.24) &  UND  &  C  &       &  ?\\ 
              &  J183656.31+382232.8  &                           &  1.15(0.04) &  2.97(0.04) &  2.53(0.08) &  BZQ  &  B  &       &  ?\\ 
              &  J183837.16+381900.5  &                           &  0.90(0.04) &  2.14(0.08) &  1.84(0.34) &  BZB  &  C  &       &  ?\\ 
J1844.3+1548  &  J184425.36+154645.9  &  NVSS J184425+154646      &  0.80(0.03) &  2.30(0.04) &  2.06(0.08) &  BZB  &  B  &  N,M  &  ?\\
J1844.9-1116  &  J184456.29-111352.1  &                           &  0.67(0.06) &  1.63(0.03) &  1.50(0.02) &  BZB  &  B  &  M    &  ?\\
J1857.6+0211  &  J185727.36+021216.4  &                           &  0.77(0.07) &  2.26(0.07) &  1.81(0.02) &  BZB  &  B  &  M,v  &  ?\\ 
J1944.3+7325  &  J194343.83+731738.7  &                           &  1.30(0.04) &  2.84(0.06) &  2.36(0.16) &  BZQ  &  C  &       &  ?\\ 
J1958.6+4020  &  J195842.28+401125.8  &                           &  1.11(0.05) &  2.82(0.06) &  2.41(0.14) &  BZQ  &  C  &       &  ?\\ 
J2042.0+4252c &  J204241.58+424150.3  &                           &  1.28(0.05) &  2.38(0.07) &  2.63(0.07) &  BZQ  &  C  &  M,v  &  ?\\
J2124.0-1513  &  J212423.63-152558.2  &                           &  1.14(0.06) &  3.15(0.11) &  2.56(0.26) &  BZQ  &  C  &       &  ?\\ 
J2128.7+5824  &  J212810.83+583336.6  &                           &  1.21(0.04) &  2.82(0.06) &  2.22(0.17) &  BZQ  &  C  &       &  ?\\ 
              &  J212820.65+582053.4  &                           &  0.54(0.03) &  1.86(0.03) &  1.78(0.10) &  BZB  &  B  &  M,v  &  ?\\
              &  J212900.37+583128.0  &                           &  1.04(0.04) &  2.37(0.13) &  2.43(0.29) &  UND  &  C  &       &  ?\\ 
J2201.2+5926  &  J215953.42+591227.1  &                           &  0.94(0.04) &  2.13(0.05) &  2.17(0.10) &  BZB  &  B  &  M    &  ?\\
\noalign{\smallskip}
\hline
\end{tabular}\\
Col. (1) 2FGL name. \\
Col. (2) \wse\ name. \\
Col. (3) Other name if present in literature and in the following order: \bzcat, NVSS, SDSS, AT20G, NED. \\
Cols. (4,5,6) Infrared colors from the \wse\ all sky catalog corrected for Galactic extinciton. Values in parentheses are 1$\sigma$ uncertainties. \\
Col. (7) Type of candidate according to our method: BZB - BZQ - UND (undetermined). \\
Col. (8) Class of candidate according to our method. \\
Col. (9) Notes: N = NVSS, F = FIRST, S = SUMSS, A=AT20G, M = 2MASS, s = SDSS dr9, 6 = 6dFGS, x = \xmm\ or \chn, X = ROSAT;  QSO  = quasar, Sy = Seyfert, LNR = LINER, BL = BL Lac; 
v = variability in \wse\ (var\_flag $>$ 5 in at least one band). \\
Col. (10) Redshift: (?) = unknown, (number?) = uncertain. 
\label{tab:ugf}
\end{table*}

\begin{table*}
\caption{Active Galaxies of Uncertain type (00h -- 12h).}
\tiny
\begin{tabular}{|lllccccclcl|}
\hline
  2FGL  &  \wse\  &  other & [3.4]-[4.6] & [4.6]-[12] & [12]-[22] & type & class & notes & z & reassoc. \\
  name  &  name   &  name  &     mag     &    mag     &   mag     &      &       &       &   & flag    \\
\hline
\noalign{\smallskip}
J0009.1+5030 &J000922.76+503028.8  &  NVSSJ000922+503028   &  0.72(0.03) &  2.19(0.05) &  1.95(0.21) &  BZB  &  C  &  M,v     &  ?       &  yes\\ 
J0018.8-8154 &J001920.58-815251.3  &  SUMSSJ001921-815251  &  0.87(0.03) &  2.30(0.03) &  1.98(0.07) &  BZB  &  B  &  M,BL    &  ?       &  yes\\
J0022.2-1853 &J002209.25-185334.7  &  NVSSJ002209-185332   &  0.87(0.03) &  2.27(0.04) &  1.83(0.12) &  BZB  &  B  &  M       &  ?       &  yes\\
J0022.3-5141 &J002200.08-514024.2  &  SUMSSJ002159-514026  &  0.81(0.03) &  2.23(0.04) &  1.87(0.14) &  BZB  &  B  &  M,v     &  ?       &  yes\\
J0045.5+1218 &J004543.33+121712.0  &  NVSSJ004543+121710   &  0.80(0.03) &  2.23(0.05) &  2.03(0.14) &  BZB  &  B  &  M       &  ?       &  yes\\
J0051.4-6241 &J005116.62-624204.3  &  SUMSSJ005116-624205  &  0.69(0.04) &  2.20(0.06) &  1.75(0.32) &  BZB  &  C  &  M       &  ?       &  yes\\
J0059.7-5700 &J005846.56-565911.4  &  SUMSSJ005846-565912  &  1.03(0.03) &  2.92(0.04) &  2.64(0.07) &  BZQ  &  B  &  M       &  ?       &  yes\\
J0110.3+6805 &J011012.84+680541.1  &  BZUJ0110+6805        &  0.67(0.03) &  2.12(0.03) &  2.02(0.04) &  BZB  &  A  &  M       &  ?       &  yes\\
J0134.4+2636 &J013428.19+263843.0  &  NVSSJ013427+263842   &  0.80(0.04) &  2.00(0.08) &  1.93(0.33) &  BZB  &  C  &  M,v     &  ?       &  yes\\
J0156.4+3909 &J015631.40+391430.9  &  NVSSJ015631+391431   &  1.13(0.04) &  2.76(0.05) &  2.47(0.11) &  BZQ  &  C  &  M,v     &  ?       &  yes\\ 
J0156.5-2419 &J015606.46-241754.3  &                       &  1.21(0.06) &  2.93(0.11) &  2.63(0.26) &  BZQ  &  C  &          &  ?       &  no \\ 
J0207.9-6832 &J020750.91-683755.1  &  SUMSSJ020750-683755  &  1.03(0.04) &  2.70(0.05) &  2.28(0.12) &  UND  &  C  &          &  ?       &  yes\\
J0210.7-5102 &J021046.19-510101.8  &  BZUJ0210-5101        &  1.13(0.03) &  2.97(0.03) &  2.39(0.04) &  BZQ  &  A  &  M       &  1.003   &  yes\\
J0238.2-3905 &J023749.42-390050.3  &                       &  0.92(0.03) &  2.40(0.04) &  2.36(0.08) &  BZB  &  C  &  M,6     &  0.323?  &  no \\
             &J023800.62-390504.6  &  NVSSJ023800-390504   &  0.68(0.04) &  2.18(0.07) &  1.91(0.27) &  BZB  &  C  &  M       &  ?       &  yes\\
J0248.6+8440 &J024948.30+843556.9  &  NVSSJ024948+843556   &  0.92(0.03) &  2.63(0.04) &  2.13(0.09) &  BZB  &  C  &  M       &  ?       &  yes\\
J0253.4+3218 &J025333.64+321720.8  &  NVSSJ025333+321721   &  1.12(0.04) &  2.87(0.06) &  2.46(0.16) &  BZQ  &  C  &  M       &  ?       &  yes\\
J0309.3-0743 &J030943.23-074427.4  &  NVSSJ030943-074427   &  0.70(0.03) &  2.15(0.06) &  2.01(0.21) &  BZB  &  C  &  M,v     &  ?       &  yes\\
J0332.5-1118 &J033200.72-111456.1  &  6dFJ0332006-111456   &  1.30(0.03) &  3.05(0.03) &  2.63(0.04) &  BZQ  &  A  &  M,6,BL  &  ?       &  no \\
             &J033223.25-111950.6  &  NVSSJ033223-111951   &  0.96(0.03) &  2.58(0.03) &  2.23(0.05) &  UND  &  C  &  M,v     &  ?       &  yes\\
J0333.7+2918 &J033349.00+291631.6  &  NVSSJ033349+291631   &  0.77(0.03) &  2.02(0.05) &  2.18(0.13) &  BZB  &  C  &  M       &  ?       &  yes\\
J0334.3+6538 &J033356.74+653656.0  &  NVSSJ033356+653656   &  0.75(0.03) &  2.07(0.05) &  1.73(0.17) &  BZB  &  C  &  M       &  ?       &  yes\\
J0424.3-5332 &J042347.22-533026.6  &                       &  1.27(0.04) &  2.86(0.04) &  2.30(0.12) &  BZQ  &  C  &  M       &  ?       &  no \\ 
             &J042504.26-533158.3  &  SUMSSJ042504-533158  &  0.99(0.03) &  2.58(0.03) &  2.12(0.04) &  BZB  &  C  &  M,v     &  ?       &  yes\\
J0433.4-6029 &J043334.08-603013.7  &  SUMSSJ043333-603014  &  1.15(0.04) &  3.04(0.04) &  2.49(0.07) &  BZQ  &  B  &  M       &  ?       &  yes\\
J0433.9-5726 &J043344.12-572613.3  &                       &  0.90(0.04) &  2.23(0.07) &  2.20(0.27) &  BZB  &  C  &          &  ?       &  no \\ 
J0438.8-4521 &J043900.84-452222.6  &  SUMSSJ043900-452223  &  1.14(0.04) &  3.09(0.04) &  2.45(0.08) &  BZQ  &  B  &          &  ?       &  yes\\
J0440.4+1433 &J044021.14+143757.0  &  NVSSJ044021+143757   &  1.13(0.05) &  2.88(0.08) &  2.54(0.20) &  BZQ  &  C  &          &  ?       &  yes\\ 
J0456.5+2658 &J045617.36+270221.1  &  NVSSJ045617+270221   &  0.95(0.04) &  2.68(0.07) &  2.35(0.17) &  UND  &  C  &          &  ?       &  yes\\
J0505.9+6116 &J050558.78+611335.9  &  NVSSJ050558+611336   &  0.65(0.04) &  1.83(0.08) &  1.69(0.45) &  BZB  &  C  &  M       &  ?       &  yes\\
J0506.7-5435 &J050657.80-543503.9  &  SUMSSJ050657-543459  &  0.74(0.03) &  1.89(0.05) &  2.02(0.19) &  BZB  &  C  &  M       &  ?       &  yes\\
J0508.1-1936 &J050805.75-194721.6  &                       &  1.11(0.04) &  2.79(0.07) &  2.72(0.13) &  BZQ  &  C  &  M       &  ?       &  no \\ 
J0512.9+4040 &J051252.53+404143.7  &  NVSSJ051252+404143   &  0.90(0.03) &  2.50(0.03) &  2.13(0.03) &  BZB  &  A  &  M       &  ?       &  yes\\
J0525.5-6011 &J052537.74-601732.0  &                       &  1.11(0.04) &  3.25(0.05) &  2.68(0.12) &  BZQ  &  C  &          &  ?       &  no \\ 
J0532.0-4826 &J053158.61-482736.0  &  SUMSSJ053158-482737  &  0.89(0.03) &  2.46(0.03) &  2.06(0.03) &  BZB  &  A  &  v       &  ?       &  yes\\
J0537.7-5716 &J053748.95-571830.2  &  SUMSSJ053748-571828  &  0.83(0.03) &  2.34(0.04) &  2.16(0.09) &  BZB  &  B  &  M       &  ?       &  yes\\
J0609.4-0248 &J060915.06-024754.6  &  NVSSJ060915-024754   &  0.78(0.04) &  1.97(0.08) &  1.71(0.37) &  BZB  &  C  &  M       &  ?       &  yes\\
J0621.9+3750 &J062157.63+375057.0  &  NVSSJ062157+375056   &  1.14(0.05) &  2.94(0.07) &  2.52(0.17) &  BZQ  &  C  &          &  ?       &  yes\\ 
J0644.2-6713 &J064428.06-671257.3  &  SUMSSJ064427-671257  &  1.06(0.03) &  2.78(0.02) &  2.26(0.03) &  UND  &  A  &          &  ?       &  yes\\
J0647.8-6102 &J064740.85-605805.2  &  SUMSSJ064740-605804  &  1.10(0.04) &  2.96(0.06) &  2.34(0.16) &  BZQ  &  C  &  v       &  ?       &  yes\\ 
             &J064806.55-610507.4  &                       &  1.38(0.04) &  3.03(0.05) &  2.48(0.12) &  BZQ  &  C  &          &  ?       &  no \\ 
J0653.7+2818 &J065344.26+281547.5  &  NVSSJ065343+281546   &  0.81(0.04) &  2.30(0.07) &  1.98(0.26) &  BZB  &  C  &  N,M     &  ?       &  no \\
J0700.3+1710 &J070001.50+170921.8  &  NVSSJ070001+170922   &  1.11(0.04) &  2.97(0.04) &  2.50(0.07) &  BZQ  &  B  &          &  ?       &  yes\\
             &J070046.29+171019.8  &                       &  1.08(0.04) &  2.62(0.08) &  2.18(0.24) &  UND  &  C  &  M       &  ?       &  no \\
J0702.7-1951 &J070242.90-195122.2  &  NVSSJ070242-195123   &  1.02(0.04) &  2.70(0.04) &  2.15(0.10) &  ?    &  B  &  M       &  ?       &  yes\\
J0703.1-3912 &J070312.64-391418.9  &  NVSSJ070312-391418   &  0.94(0.03) &  2.51(0.04) &  2.15(0.09) &  BZB  &  B  &  M       &  ?       &  yes\\
J0706.5+7741 &J070651.32+774137.0  &  NVSSJ070651+774137   &  0.90(0.03) &  2.46(0.03) &  1.91(0.08) &  BZB  &  B  &  M       &  ?       &  yes\\
J0706.7-4845 &J070549.12-483911.4  &                       &  0.95(0.04) &  2.86(0.06) &  2.38(0.19) &  UND  &  C  &  M       &  ?       &  no \\ 
J0709.3-0256 &J070945.05-025517.4  &  NVSSJ070945-025517   &  1.09(0.03) &  2.91(0.03) &  2.30(0.06) &  UND  &  C  &  v       &  ?       &  yes\\
J0726.0-0053 &J072550.63-005456.4  &  BZUJ0725-0054        &  1.08(0.03) &  2.98(0.03) &  2.45(0.03) &  BZQ  &  A  &  M       &  0.128   &  yes\\
J0734.2-7706 &J073443.44-771113.4  &  SUMSSJ073441-771113  &  1.04(0.04) &  3.20(0.05) &  2.48(0.11) &  BZQ  &  C  &          &  ?       &  yes\\ 
J0746.5-0713 &J074627.48-070949.7  &  NVSSJ074627-070951   &  1.10(0.04) &  3.04(0.07) &  2.33(0.18) &  UND  &  C  &  M       &  ?       &  yes\\
J0746.5-4758 &J074642.30-475455.2  &  SUMSSJ074642-475455  &  1.15(0.03) &  2.91(0.06) &  1.89(0.16) &  BZB  &  C  &  M       &  ?       &  yes\\
J0816.7-2420 &J081639.46-242635.4  &                       &  1.13(0.07) &  3.20(0.11) &  2.59(0.30) &  BZQ  &  C  &          &  ?       &  no \\ 
             &J081640.41-242106.6  &  NVSSJ081640-242105   &  1.10(0.04) &  3.04(0.04) &  2.45(0.09) &  BZQ  &  B  &          &  ?       &  yes\\
J0823.0+4041 &J082257.55+404149.8  &  NVSSJ082257+404149   &  1.15(0.04) &  2.91(0.04) &  2.32(0.07) &  BZQ  &  C  & QSO      &  0.8657 &  yes\\
J0844.8-5459 &J084502.47-545808.5  &  AT20GJ084502-545808  &  1.02(0.03) &  2.82(0.03) &  2.31(0.04) &  UND  &  B  &  M       &  ?       &  yes\\
J0849.9-3540 &J084945.61-354101.2  &  NVSSJ084945-354102   &  1.01(0.03) &  2.53(0.04) &  2.35(0.08) &  UND  &  B  &  v       &  ?       &  yes\\
J0852.4-5756 &J085238.73-575529.4  &  AT20GJ085238-575530  &  1.23(0.03) &  2.87(0.03) &  2.38(0.03) &  BZQ  &  A  &          &  ?       &  yes\\
J0853.1-3659 &J085310.50-365823.1  &  NVSSJ085310-365820   &  0.77(0.04) &  2.32(0.04) &  1.98(0.08) &  BZB  &  B  &  N,M     &  ?       &  no \\
J0855.1-0712 &J085435.20-071837.5  &                       &  1.10(0.03) &  3.03(0.03) &  2.48(0.05) &  BZQ  &  B  &  M       &  ?       &  no \\
J0856.0+7136 &J085654.85+714623.8  &  NVSSJ085654+714624   &  1.09(0.04) &  2.97(0.04) &  2.49(0.07) &  BZQ  &  B  &  M       &  ?       &  yes\\
J0858.1-1952 &J085805.36-195036.8  &  NVSSJ085805-195036   &  1.16(0.04) &  2.80(0.05) &  2.25(0.11) &  UND  &  C  &  M,QSO   &  0.6597  &  yes\\
J0904.8-3513 &J090423.42-351203.0  &                       &  1.24(0.06) &  2.82(0.10) &  2.37(0.32) &  BZQ  &  C  &          &  ?       &  no \\ 
J0906.2-0906 &J090618.05-090544.9  &  NVSSJ090618-090544   &  1.01(0.04) &  2.56(0.05) &  2.36(0.12) &  UND  &  C  &  M       &  ?       &  yes\\
J0919.3-2203 &J092002.74-215835.0  &                       &  1.01(0.04) &  2.76(0.06) &  2.21(0.19) &  UND  &  C  &  M       &  ?       &  no \\ 
J0940.8-6105 &J094047.33-610728.5  &  AT20GJ094047-610726  &  0.87(0.03) &  2.57(0.03) &  2.17(0.06) &  BZB  &  B  &  M,v     &  ?       &  yes\\
J0941.9-0755 &J094221.46-075953.1  &  NVSSJ094221-075953   &  0.95(0.03) &  2.50(0.03) &  2.21(0.07) &  BZB  &  B  &  v       &  ?       &  yes\\
J0946.9-2541 &J094709.52-254100.0  &  NVSSJ094709-254100   &  0.72(0.04) &  2.16(0.06) &  2.06(0.23) &  BZB  &  C  &  M       &  ?       &  yes\\
J1016.2-0638 &J101542.96-063055.1  &                       &  0.93(0.04) &  2.91(0.07) &  2.48(0.16) &  UND  &  C  &          &  ?       &  no \\ 
             &J101626.98-063625.2  &  NVSSJ101626-06362    &  0.91(0.04) &  2.88(0.07) &  2.38(0.18) &  UND  &  C  &  N,F     &  ?       &  no \\
J1045.5-2931 &J104540.62-292726.4  &  NVSSJ104540-292725   &  1.09(0.05) &  3.13(0.06) &  2.33(0.17) &  BZQ  &  C  &          &  ?       &  yes\\ 
J1103.9-5356 &J110352.22-535700.7  &  AT20GJ110352-535700  &  0.98(0.03) &  2.88(0.03) &  2.22(0.03) &  ?    &  A  &  M,v     &  ?       &  yes\\
J1106.3-3643 &J110624.04-364659.0  &  NVSSJ110624-364659   &  1.08(0.04) &  2.65(0.06) &  2.39(0.16) &  ?    &  C  &          &  ?       &  yes\\
J1154.1-3242 &J115406.16-324243.0  &  NVSSJ115406-324242   &  1.02(0.04) &  2.87(0.05) &  2.41(0.12) &  ?    &  C  &  M,v     &  ?       &  yes\\
J1230.2-5258 &J122939.88-530332.1  &  AT20GJ122939-530332  &  0.66(0.04) &  2.22(0.07) &  1.90(0.27) &  BZB  &  C  &  M       &  ?       &  no \\
J1238.1-1953 &J123824.40-195913.4  &  NVSSJ123824-195913   &  0.88(0.03) &  2.46(0.04) &  2.13(0.11) &  BZB  &  B  &  M       &  ?       &  yes\\
J1239.5+0728 &J123924.58+073017.2  &  BZUJ1239+0730        &  1.07(0.03) &  2.96(0.04) &  2.32(0.07) &  ?    &  C  &  M       &  0.4     &  no \\
\noalign{\smallskip}
\hline
\end{tabular}\\
Col. (1) 2FGL name. \\
Col. (2) \wse\ name. \\
Col. (3) Other name if present in literature and in the following order: \bzcat, NVSS, SDSS, AT20G, NED. \\
Cols. (4,5,6) Infrared colors from the \wse\ all sky catalog corrected for Galactic extinciton. Values in parentheses are 1$\sigma$ uncertainties. \\
Col. (7) Type of candidate according to our method: BZB - BZQ - UND (undetermined). \\
Col. (8) Class of candidate according to our method. \\
Col. (9) Notes: N = NVSS, F = FIRST, S = SUMSS, A=AT20G, M = 2MASS, s = SDSS dr9, 6 = 6dFGS, x = \xmm\ or \chn, X = ROSAT;  QSO  = quasar, Sy = Seyfert, LNR = LINER, BL = BL Lac; 
v = variability in \wse\ (var\_flag $>$ 5 in at least one band). \\
Col. (10) Redshift: (?) = unknown, (number?) = uncertain. 
Col. (11) Re-association flag: ``yes'' if the association of our method corresponds to the one provided in the 2FGL, ``no'' otherwise.
\label{tab:agu1}
\end{table*}

\begin{table*}
\caption{Active galaxies of Uncertain type (12h -- 24h).}
\tiny
\begin{tabular}{|lllccccclcl|}
\hline
  2FGL  &  \wse\  &  other & [3.4]-[4.6] & [4.6]-[12] & [12]-[22] & type & class & notes & z & reassoc. \\
  name  &  name   &  name  &     mag     &    mag     &   mag     &      &       &       &   & flag    \\
\hline
\noalign{\smallskip}
J1301.6+3331 &J130147.03+332236.3  &SDSSJ130147.01+332236.4&  1.04(0.03) &  2.46(0.04) &  2.27(0.08) &  UND  &  B  &  M,s     &  ?       &  no \\
J1303.8-5537 &J130349.23-554031.6  &  AT20GJ130349-554031  &  1.01(0.03) &  2.87(0.03) &  2.35(0.03) &  UND  &  A  &  M,v     &  ?       &  yes\\
J1304.1-2415 &J130416.70-241216.6  &  NVSSJ130416-24121    &  0.89(0.03) &  2.37(0.05) &  2.01(0.15) &  BZB  &  C  &  N,M     &  ?       &  no \\
J1304.3-4353 &J130421.01-435310.2  &  SUMSSJ130420-435308  &  0.84(0.03) &  2.24(0.03) &  1.93(0.05) &  BZB  &  B  &  M,v     &  ?       &  yes\\
J1307.5-4300 &J130737.98-425938.9  &  SUMSSJ130737-42594   &  0.73(0.03) &  2.04(0.03) &  1.90(0.08) &  BZB  &  B  &  S,M,6,v &  ?       &  no \\
J1307.6-6704 &J130817.51-670705.8  &  AT20GJ130817-670704  &  0.77(0.03) &  2.55(0.04) &  2.28(0.06) &  BZB  &  B  &  M,v     &  ?       &  yes\\
J1329.2-5608 &J132901.16-560802.5  &  AT20GJ132901-560802  &  1.05(0.03) &  2.78(0.03) &  2.22(0.07) &  UND  &  B  &  M,v     &  ?       &  yes\\
J1330.1-7002 &J133011.34-700312.7  &  AT20GJ133010-700313  &  0.78(0.03) &  2.48(0.03) &  2.25(0.04) &  BZB  &  B  &          &  ?       &  yes\\
J1351.3-2909 &J135146.85-291217.4  &  BZUJ1351-2912        &  1.12(0.03) &  2.87(0.03) &  2.48(0.05) &  BZQ  &  B  &  v       &  1.034?  &  no \\
J1406.2-2510 &J140609.60-250809.2  &  NVSSJ140609-250808   &  0.83(0.04) &  2.37(0.07) &  2.02(0.25) &  BZB  &  C  &  M,v     &  ?       &  yes\\
J1416.3-2415 &J141554.91-241925.8  &  NVSSJ141554-241924   &  1.26(0.05) &  3.13(0.08) &  2.51(0.22) &  BZQ  &  C  &  N       &  ?       &  no \\ 
             &J141642.23-241021.2  &                       &  0.92(0.04) &  2.71(0.08) &  2.39(0.23) &  UND  &  C  &  M       &  ?       &  no \\ 
J1419.4-0835 &J141922.56-083831.9  &  NVSSJ141922-083830   &  1.01(0.03) &  2.78(0.04) &  2.15(0.09) &  UND  &  B  &  v       &  ?       &  yes\\
J1514.6-4751 &J151440.03-474829.7  &  AT20GJ151440-474828  &  1.14(0.04) &  3.07(0.04) &  2.43(0.06) &  BZQ  &  B  &  M       &  ?       &  yes\\
J1518.2-2733 &J151803.60-273131.0  &  NVSSJ151803-273131   &  0.62(0.03) &  2.14(0.04) &  2.18(0.10) &  BZB  &  C  &  M       &  ?       &  yes\\
J1553.2-2424 &J155331.62-242206.0  &  NVSSJ155331-242206   &  1.16(0.04) &  2.95(0.07) &  2.66(0.14) &  BZQ  &  C  &          &  0.332?  &  yes\\ 
J1558.3+8513 &J160031.76+850949.2  &  NVSSJ160031+850948   &  0.82(0.03) &  2.18(0.04) &  2.02(0.14) &  BZB  &  B  &  N,M     &  ?       &  no \\
J1604.5-4442 &J160431.03-444131.9  &  AT20GJ160431-444131  &  0.98(0.03) &  2.80(0.03) &  2.46(0.05) &  UND  &  B  &  M       &  ?       &  yes\\
J1626.0-7636 &J162638.17-763855.4  &  SUMSSJ162639-763856  &  0.56(0.03) &  2.21(0.05) &  1.97(0.19) &  BZB  &  C  &  M,v     &  ?       &  yes\\
J1650.1-5044 &J165016.63-504448.2  &  AT20GJ165016-504446  &  0.95(0.04) &  2.65(0.04) &  2.22(0.09) &  UND  &  B  &  M,v     &  ?       &  yes\\
J1725.1-7714 &J172350.86-771350.3  &  SUMSSJ172350-771350  &  1.01(0.03) &  2.67(0.04) &  2.33(0.08) &  UND  &  B  &  M,v     &  ?       &  yes\\
J1759.2-4819 &J175858.45-482112.4  &  SUMSSJ175858-482112  &  0.91(0.03) &  2.62(0.03) &  2.21(0.05) &  UND  &  C  &          &  ?       &  yes\\
J1811.0+5340 &J181037.99+533501.5  &  NVSSJ181038+533501   &  0.83(0.03) &  2.22(0.05) &  2.15(0.15) &  BZB  &  C  &  N,M,X   &  ?       &  no \\
J1815.6-6407 &J181425.96-641008.8  &                       &  1.27(0.06) &  3.05(0.08) &  2.67(0.19) &  BZQ  &  C  &          &  ?       &  no \\ 
J1816.7-4942 &J181655.99-494344.7  &  SUMSSJ181655-494344  &  0.99(0.05) &  2.86(0.07) &  2.21(0.19) &  UND  &  C  &          &  ?       &  yes\\ 
J1818.7+2138 &J181905.22+213234.0  &  NVSSJ181905+213235   &  0.90(0.04) &  2.50(0.05) &  1.99(0.17) &  BZB  &  C  &  M       &  ?       &  yes\\
J1820.6+3625 &J182023.61+362914.4  &                       &  1.04(0.04) &  2.61(0.05) &  2.36(0.17) &  UND  &  C  &          &  ?       &  no \\
J1823.6-3453 &J182338.59-345412.0  &  NVSSJ182338-345412   &  0.68(0.04) &  2.05(0.04) &  1.92(0.13) &  BZB  &  B  &  M       &  ?       &  yes\\
J1823.8+4312 &J182352.33+431452.5  &                       &  1.12(0.04) &  2.96(0.05) &  2.50(0.13) &  BZQ  &  C  &  x       &  ?       &  no \\ 
J1825.1-5231 &J182513.79-523058.1  &  SUMSSJ182513-523057  &  1.65(0.03) &  2.86(0.03) &  2.26(0.07) &  UND  &  B  &  M       &  ?       &  yes\\
J1830.0+1325 &J183000.76+132414.4  &  NVSSJ183000+132414   &  0.95(0.04) &  2.60(0.04) &  2.23(0.11) &  UND  &  C  &  M       &  ?       &  yes\\
J1830.2-4441 &J183000.86-444111.4  &  SUMSSJ183000-444112  &  1.17(0.04) &  3.10(0.05) &  2.37(0.10) &  BZQ  &  C  &  M       &  ?       &  yes\\
J1844.7+5716 &J184450.96+570938.6  &  NVSSJ184451+570940   &  0.86(0.03) &  2.38(0.04) &  2.16(0.10) &  BZB  &  B  &  M       &  ?       &  yes\\
J1936.9+8402 &J193930.23+835925.8  &                       &  1.08(0.04) &  2.97(0.05) &  2.43(0.14) &  BZQ  &  C  &          &  ?       &  no \\ 
J1940.8-6213 &J194121.76-621120.8  &  SUMSSJ194121-621120  &  1.30(0.04) &  2.88(0.04) &  2.36(0.09) &  BZQ  &  C  &          &  ?       &  yes\\
J1942.8+1033 &J194247.48+103327.2  &  NVSSJ194247+103327   &  0.72(0.03) &  2.06(0.03) &  1.85(0.07) &  BZB  &  B  &  M       &  ?       &  yes\\
J1959.9-4727 &J195945.66-472519.2  &  SUMSSJ195945-472519  &  0.80(0.03) &  2.15(0.04) &  1.66(0.17) &  BZB  &  C  &  M       &  ?       &  yes\\
J2040.2-7109 &J203931.44-711033.0  &                       &  1.12(0.04) &  3.01(0.05) &  2.71(0.12) &  BZQ  &  C  &          &  ?       &  no \\ 
J2049.8+1001 &J204932.28+095911.8  &                       &  1.23(0.06) &  3.13(0.10) &  2.68(0.23) &  BZQ  &  C  &          &  ?       &  no \\ 
J2103.6-6236 &J210338.38-623225.8  &  SUMSSJ210338-623226  &  0.81(0.03) &  2.30(0.03) &  1.80(0.12) &  BZB  &  B  &  M       &  ?       &  yes\\
J2250.2-4205 &J225014.94-420218.6  &                       &  1.05(0.04) &  2.64(0.07) &  2.39(0.20) &  UND  &  C  &          &  ?       &  no \\ 
             &J225022.20-420613.2  &  SUMSSJ225022-420613  &  0.94(0.03) &  2.47(0.04) &  1.88(0.13) &  BZB  &  C  &  M       &  0.1187? &  yes\\ 
J2317.3-4534 &J231731.97-453359.6  &  SUMSSJ231731-453400  &  0.70(0.04) &  2.19(0.08) &  1.92(0.34) &  BZB  &  C  &  S,M,6   &  ?       &  no \\
J2323.0-4918 &J232255.30-491942.0  &                       &  1.08(0.06) &  3.21(0.10) &  2.34(0.30) &  BZQ  &  C  &          &  ?       &  no \\ 
J2324.6+0801 &J232445.31+080206.3  &  NVSSJ232445+080205   &  0.83(0.04) &  2.30(0.06) &  2.05(0.21) &  BZB  &  C  &  M       &  ?       &  yes\\
J2325.4+1650 &J232526.62+164941.1  &                       &  1.34(0.04) &  3.04(0.05) &  2.42(0.12) &  BZQ  &  C  &          &  ?       &  no \\ 
             &J232538.11+164642.8  &  NVSSJ232538+164641   &  0.78(0.04) &  2.15(0.10) &  2.27(0.33) &  BZB  &  C  &  M       &  ?       &  yes\\
\noalign{\smallskip}
\hline
\end{tabular}\\
Col. (1) 2FGL name. \\
Col. (2) \wse\ name. \\
Col. (3) Other name if present in literature and in the following order: \bzcat, NVSS, SDSS, AT20G, NED. \\
Cols. (4,5,6) Infrared colors from the \wse\ all sky catalog corrected for Galactic extinciton. Values in parentheses are 1$\sigma$ uncertainties. \\
Col. (7) Type of candidate according to our method: BZB - BZQ - UND (undetermined). \\
Col. (8) Class of candidate according to our method. \\
Col. (9) Notes: N = NVSS, F = FIRST, S = SUMSS, A=AT20G, M = 2MASS, s = SDSS dr9, 6 = 6dFGS, x = \xmm\ or \chn, X = ROSAT;  QSO  = quasar, Sy = Seyfert, LNR = LINER, BL = BL Lac; 
v = variability in \wse\ (var\_flag $>$ 5 in at least one band). \\
Col. (10) Redshift: (?) = unknown, (number?) = uncertain. 
Col. (11) Re-association flag: ``yes'' if the association of our method corresponds to the one provided in the 2FGL, ``no'' otherwise.
\label{tab:agu2}
\end{table*}

\begin{table}
\caption{UGSs without $\gamma$-ray blazar candidates associated.}
\begin{tabular}{|lll|}
\hline
\noalign{\smallskip}
2FGLJ0002.7+6220 &   2FGLJ1115.0-0701 &   2FGLJ1721.0+0711\\
2FGLJ0032.7-5521 &   2FGLJ1117.2-5341 &   2FGLJ1722.5-0420\\
2FGLJ0048.8-6347 &   2FGLJ1120.0-2204 &   2FGLJ1730.6-2409\\
2FGLJ0212.1+5318 &   2FGLJ1129.0-0532 &   2FGLJ1744.1-7620\\
2FGLJ0224.0+6204 &   2FGLJ1208.5-6240 &   2FGLJ1747.6+0324\\
2FGLJ0239.5+1324 &   2FGLJ1221.4-0633 &   2FGLJ1748.8+3418\\
2FGLJ0248.5+5131 &   2FGLJ1231.3-5112 &   2FGLJ1748.9-3923\\
2FGLJ0305.0-1602 &   2FGLJ1240.6-7151 &   2FGLJ1753.8-4446\\
2FGLJ0307.4+4915 &   2FGLJ1306.2-6044 &   2FGLJ1757.5-6028\\
2FGLJ0318.0+0255 &   2FGLJ1312.9-2351 &   2FGLJ1759.4-2954\\
2FGLJ0336.0+7504 &   2FGLJ1335.3-4058 &   2FGLJ1820.6-3219\\
2FGLJ0338.2+1306 &   2FGLJ1353.5-6640 &   2FGLJ1821.8+0830\\
2FGLJ0353.2+5653 &   2FGLJ1400.7-1438 &   2FGLJ1828.7+3231\\
2FGLJ0359.5+5410 &   2FGLJ1410.4+7411 &   2FGLJ1830.9-3132\\
2FGLJ0409.5+0509 &   2FGLJ1417.7-5028 &   2FGLJ1902.7-7053\\
2FGLJ0418.9+6636 &   2FGLJ1423.9-7842 &   2FGLJ1906.5+0720\\
2FGLJ0420.9-3743 &   2FGLJ1424.2-1752 &   2FGLJ1919.5-7324\\
2FGLJ0426.7+5434 &   2FGLJ1458.5-2121 &   2FGLJ1946.7-1118\\
2FGLJ0438.0-7331 &   2FGLJ1507.0-6223 &   2FGLJ1947.8-0739\\
2FGLJ0439.8-1858 &   2FGLJ1513.5-2546 &   2FGLJ2002.8-2150\\
2FGLJ0516.7+2634 &   2FGLJ1513.9-2256 &   2FGLJ2017.5-1618\\
2FGLJ0523.3-2530 &   2FGLJ1518.4-5233 &   2FGLJ2018.0+3626\\
2FGLJ0524.1+2843 &   2FGLJ1536.4-4949 &   2FGLJ2034.7-4201\\
2FGLJ0533.9+6759 &   2FGLJ1539.2-3325 &   2FGLJ2034.9+3632\\
2FGLJ0539.3-0323 &   2FGLJ1544.5-1126 &   2FGLJ2041.2+4735\\
2FGLJ0605.3+3758 &   2FGLJ1548.3+1453 &   2FGLJ2042.8-7317\\
2FGLJ0658.4+0633 &   2FGLJ1601.1-4220 &   2FGLJ2046.2-4259\\
2FGLJ0719.2-5000 &   2FGLJ1617.3-5336 &   2FGLJ2103.5-1112\\
2FGLJ0758.8-1448 &   2FGLJ1617.5-2657 &   2FGLJ2107.8+3652\\
2FGLJ0803.2-0339 &   2FGLJ1622.8-5006 &   2FGLJ2110.3+3822\\
2FGLJ0843.6+6715 &   2FGLJ1624.1-4040 &   2FGLJ2112.5-3042\\
2FGLJ0854.7-4501 &   2FGLJ1626.4-4408 &   2FGLJ2115.4+1213\\
2FGLJ0859.4-2532 &   2FGLJ1631.6-2819 &   2FGLJ2117.5+3730\\
2FGLJ0934.0-6231 &   2FGLJ1646.7-1333 &   2FGLJ2212.6+0702\\
2FGLJ0952.7-3717 &   2FGLJ1649.2-3004 &   2FGLJ2246.3+1549\\
2FGLJ1016.4-4244 &   2FGLJ1704.3+1235 &   2FGLJ2249.1+5758\\
2FGLJ1033.5-5032 &   2FGLJ1704.6-0529 &   2FGLJ2339.6-0532\\
2FGLJ1036.1-6722 &   2FGLJ1709.0-0821 &   2FGLJ2351.6-7558\\
\noalign{\smallskip}
\hline
\end{tabular}
\label{tab:dm}
\end{table}

\section{Comparison with other methods}
\label{sec:comparison}
Among the whole sample of 590 UGSs analyzed, 299 without and 291 with $\gamma$-ray analysis flag 
there are 28 sources having at least one $\gamma$-ray blazar candidate
that were also unidentified in the First Fermi $\gamma$-ray LAT catalog \citep[1FGL;][]{abdo10} and { they}
were analyzed using two different statistical approaches: the Classification Tree
and the Logistic regression analyses \citep[see][and references therein]{ackermann12}.
For these 28 UGSs, analyzed on the basis of the above statistical approaches, 
we performed a comparison with our results to verify if the 2FGL sources that we associated with a $\gamma$-ray blazar 
candidates have been also classified as AGNs.

By comparing the results of our association method with those in Ackermann et al. (2012), we found that
23 out of 28 UGSs that we associate with $\gamma$-ray blazar candidates are classified as AGNs,
all of them with a probability higher than 66\% and 12 of them higher than 80\% \citep[see][]{ackermann12}. 
Among the remaining 5 sources, 4 have been classified as pulsars,
with a very low probability with respect to the whole sample, systematically lower than 56\%.
In addition, there is one with an ambiguous classification.
Consequently, we emphasize that { for the subamples where we overlap} our results are in good agreement with the classification 
suggested by Ackermann et al. (2012), consistent with the $\gamma$-ray blazar nature of the \wse\ candidates proposed in our analysis.

\section{Summary and conclusions}
\label{sec:summary}
A new association method has been recently developed on the basis of the striking discovery that 
$\gamma$-ray emitting blazars occupy a distinct region in the \wse\ 3-dimensional color space,
{ well separated} from that occupied by other extragalactic and galactic sources \citep{paper1,paper2}.
According to D'Abrusco et al. (2013) the 3-dimensional region occupied 
by $\gamma$-ray emitting blazars { is} the $locus$;
its 2-dimensional projection in the [3.4]-[4.6]-[12] $\mu$m parameter space,
{ retains} its historical definition of \wse\ Gamma-ray Strip \citep{paper1}.
Additional improvements, mostly based on the \wse\ all-sky data release, available since March 2012 \citep[e.g.,][]{cutri12},
and on a new parametrization of the $locus$ in the parameter space 
of its principal components have been subsequently developed \citep{paper6}.

%
In this work we { describe} the results obtained by applying our new association procedure
to { the} search for new $\gamma$-ray blazar candidates in the two samples: 
the unidentified gamma-ray sources (UGSs), and the active galaxies 
of uncertain type (AGUs), { as} listed in the 2FGL \citep{nolan12}. 

We present the complete list of $\gamma$-ray blazar candidates found using the \wse\ observations.
We also perform an extensive archival search to see if the sources associated with our method,
show additional blazar-like characteristics; as { for example} the presence of a radio counterpart and/or of a spectrum that could be 
featureless as for BZBs or similar to those of broad-line quasars as generally occurs in BZQs.

We found 62 $\gamma$-ray blazar candidates for the UGS without any $\gamma$-ray analysis flag
and 49 for those with $\gamma$-ray analysis flag, out of a total of 590 sources investigated.
For the AGUs sample, we confirmed the blazar-like nature of 87 out 210 of AGUs 
analyzed on the basis of their IR colors.

A significant fraction (i.e., $\sim$ 36\%) of the \wse\ sources associated with our method { with} UGSs
have a radio counterpart, more than 50\% are also detected in the 2MASS catalog as generally occurs
for blazars, and more than $\sim$10\% appear to be variable according to the \wse\ analysis flags \citep{cutri12}.
Notably, all the sources for which an optical spectrum was available in literature 
clearly show blazar-like features, being either featureless or having broad emission lines typical of quasars, 
the only exception being SDSS J015910.05+010514.5, one of the counterparts
associated with 2FGLJ0158.4+0107.
As generally expected for $\gamma$-ray blazars a handful of the selected candidates are also detected in the X-rays. 
A deeper investigation of their X-ray counterparts will be addressed in a forthcoming paper \citep{paggi13}.
All the $\gamma$-ray blazar candidates selected with our association procedure appear to be extragalactic in nature; moreover
our selection seems not to be highly contaminated by any class of non-blazar-like sources, as for example obscured quasars 
or Seyfert galaxies.

{ Our results are in good agreement with those based on different statistical approaches
like the Classification Tree and the Logistic regression analyses \citep{ackermann12}.
In particular, 23 out of 28 UGSs that we associate to a $\gamma$-ray blazar candidate are also classified as active galaxies
by the above methods at high level of confidence.}

For UGSs associated with a pulsar in the 2FGL analysis as reported in the 
Public List of LAT-Detected Gamma-Ray Pulsars (see Section~\ref{sec:ugs}),
we did not find any \wse\ $\gamma$-ray blazar candidate, confirming the { reliability} of our selection procedure.
We provide a list of the UGSs for which we did not find any $\gamma$-ray blazar candidates
using { either} the new improved method { or} the old parametrization (i.e., less conservative),
within their positional uncertainty regions at 95\% level of confidence.
This list of \fer\ sources reported in Table~\ref{tab:dm} could be useful for follow up observations aiming at discover new pulsars
or to constrain exotic high-energy physics phenomena such as dark matter signatures, or new classes of sources \citep[e.g., ][]{zechlin12,su12}.

Finally, we emphasize that { additional} investigations of different samples of active galactic nuclei, such as Seyfert galaxies, 
are necessary to study the problem of the contamination of our association method by 
extragalactic sources with infrared colors similar to those of $\gamma$-ray blazars. 
Moreover extensive ground-based spectroscopic follow up observations in the optical and in the near IR
{ would be ideal} to verify the nature of the selected \wse\ sources and to estimate the fraction of non-blazar objects,
{ similar to the recent studies} performed for the unidentified INTEGRAL sources \citep[e.g.,][]{masetti08,masetti09,masetti10,masetti12}.
\vspace{0.5cm}

{\it Note added to the proofs:} The infrared source WISE J182352.33+431452.5, potential counterpart
of 2FGL J1823.8+4312, is a possible contaminant of our selections given its optical spectrum typical of an
obscured red quasar (D. Stern priv. comm.).

\acknowledgements
We thank the anonymous referee for useful comments that led to improvements in the paper.
F. Massaro is grateful to S. Digel and D. Thompson for their helpful discussions
and to M. Ajello, E. Ferrara and J. Ballet for their support.
The work is supported by the NASA grants NNX12AO97G.
R. D'Abrusco gratefully acknowledges the financial 
support of the US Virtual Astronomical Observatory, which is sponsored by the
National Science Foundation and the National Aeronautics and Space Administration.
The work by G. Tosti is supported by the ASI/INAF contract I/005/12/0.
H. A. Smith acknowledges partial support from NASA/JPL grant RSA 1369566.
TOPCAT\footnote{\underline{http://www.star.bris.ac.uk/$\sim$mbt/topcat/}} 
\citep{taylor2005} and SAOImage DS9 were used extensively in this work 
for the preparation and manipulation of the tabular data and the images.
Part of this work is based on archival data, software or on-line services provided by the ASI Science Data Center.
This research has made use of data obtained from the High Energy Astrophysics Science Archive
Research Center (HEASARC) provided by NASA's Goddard
Space Flight Center; the SIMBAD database operated at CDS,
Strasbourg, France; the NASA/IPAC Extragalactic Database
(NED) operated by the Jet Propulsion Laboratory, California
Institute of Technology, under contract with the National Aeronautics and Space Administration.
Part of this work is based on the NVSS (NRAO VLA Sky Survey);
The National Radio Astronomy Observatory is operated by Associated Universities,
Inc., under contract with the National Science Foundation. 
This publication makes use of data products from the Two Micron All Sky Survey, which is a joint project of the University of 
Massachusetts and the Infrared Processing and Analysis Center/California Institute of Technology, funded by the National Aeronautics 
and Space Administration and the National Science Foundation.
This publication makes use of data products from the Wide-field Infrared Survey Explorer, 
which is a joint project of the University of California, Los Angeles, and 
the Jet Propulsion Laboratory/California Institute of Technology, 
funded by the National Aeronautics and Space Administration.
Funding for the SDSS and SDSS-II has been provided by the Alfred P. Sloan Foundation, 
the Participating Institutions, the National Science Foundation, the U.S. Department of Energy, 
the National Aeronautics and Space Administration, the Japanese Monbukagakusho, 
the Max Planck Society, and the Higher Education Funding Council for England. 
The SDSS Web Site is http://www.sdss.org/.
The SDSS is managed by the Astrophysical Research Consortium for the Participating Institutions. 
The Participating Institutions are the American Museum of Natural History, 
Astrophysical Institute Potsdam, University of Basel, University of Cambridge, 
Case Western Reserve University, University of Chicago, Drexel University, 
Fermilab, the Institute for Advanced Study, the Japan Participation Group, 
Johns Hopkins University, the Joint Institute for Nuclear Astrophysics, 
the Kavli Institute for Particle Astrophysics and Cosmology, the Korean Scientist Group, 
the Chinese Academy of Sciences (LAMOST), Los Alamos National Laboratory, 
the Max-Planck-Institute for Astronomy (MPIA), the Max-Planck-Institute for Astrophysics (MPA), 
New Mexico State University, Ohio State University, University of Pittsburgh, 
University of Portsmouth, Princeton University, the United States Naval Observatory, 
and the University of Washington.

{}

\appendix
\section{Optical counterparts}
\label{sec:appendix}
In Tables~\ref{tab:ugs_usno}, \ref{tab:ugf_usno}, \ref{tab:agu_usno1} and \ref{tab:agu_usno2}, we report the magnitudes 
of the optical counterpart uniquely found within 3\arcsec.3,
for all the $\gamma$-ray blazar candidates, selected according to our association procedure.
This information permits { us} to optimize the strategy for { the} future follow up optical observations needed to clarify the nature of the 
selected sources and to determine their redshifts via spectroscopy.

\begin{table}
\caption{Optical magnitudes of the USNO B1 catalog for the UGSs without $\gamma$-ray analysis flags.}
\tiny
\begin{tabular}{|llccccc|}
\hline
  2FGL   &   \wse\   &   B1     &   R1     &   B2     &   R2     &   I    \\ 
  name   &   name    &   mag    &   mag    &   mag    &   mag    &   mag   \\ 
\hline
\noalign{\smallskip}
J0039.1+4331  &  J003858.27+432947.0  &  19.29  &  19.26  &  19.03  &  18.84  &  18.32\\
J0116.6-6153  &  J011619.59-615343.5  &         &  17.72  &  18.22  &  17.78  &  17.91\\
J0133.4-4408  &  J013306.35-441421.3  &         &  18.38  &  19.70  &  18.12  &  18.76\\
J0133.4-4408  &  J013321.36-441319.4  &         &  18.63  &  19.39  &  18.32  &  18.80\\
J0143.6-5844  &  J014347.39-584551.3  &         &  16.70  &  18.48  &  16.64  &  17.04\\
J0158.4+0107  &  J015910.05+010514.7  &  17.94  &  18.27  &  17.71  &  18.43  &  16.33\\
J0158.4+0107  &  J015757.45+011547.8  &  21.13  &  19.39  &         &  19.31  &  18.58\\
J0158.4+0107  &  J015836.25+010632.1  &  19.22  &  18.90  &  19.02  &  19.07  &  17.62\\
J0158.6+8558  &  J014935.30+860115.3  &  18.79  &  17.14  &  18.60  &  16.96  &  15.92\\
J0158.6+8558  &  J015550.16+854745.1  &  19.83  &  19.27  &  19.95  &  18.90  &  17.78\\
J0227.7+2249  &  J022744.35+224834.3  &         &         &  20.82  &  20.22  &  19.28\\
J0316.1-6434  &  J031614.31-643731.4  &         &  16.59  &  18.19  &  16.57  &  16.82\\
J0332.1+6309  &  J033153.90+630814.1  &         &         &  20.66  &  19.92  &  18.35\\
J0409.8-0357  &  J040946.57-040003.4  &  19.45  &  19.18  &  17.53  &  16.98  &  16.86\\
J0414.9-0855  &  J041457.01-085652.0  &  20.48  &  18.15  &  20.21  &  17.31  &  17.69\\
J0416.0-4355  &  J041605.81-435514.6  &         &  18.49  &  18.70  &  18.17  &  18.00\\
J0555.9-4348  &  J055618.74-435146.1  &         &  19.23  &  18.88  &  19.08  &  18.08\\
J0555.9-4348  &  J055531.59-435030.7  &         &  19.16  &  19.22  &  18.93  &  18.19\\
J0602.7-4011  &  J060237.10-401453.2  &         &  17.61  &  17.76  &  17.74  &  17.62\\
J0644.6+6034  &  J064459.38+603131.7  &  19.44  &  19.03  &  19.33  &  18.23  &  18.28\\
J0713.5-0952  &  J071223.28-094536.3  &         &         &  19.48  &  17.88  &  17.24\\
J0723.9+2901  &  J072354.83+285929.9  &  19.78  &  19.05  &  19.97  &  18.72  &       \\
J0744.1-2523  &  J074402.19-252146.0  &         &  13.33  &         &  9.510  &       \\
J0746.0-0222  &  J074627.03-022549.3  &  19.03  &         &  18.59  &  18.43  &  16.53\\
J0756.3-6433  &  J075624.60-643030.6  &         &  18.80  &  19.13  &  17.26  &  18.56\\
J0807.0-6511  &  J080729.66-650910.3  &         &  18.97  &  19.15  &  19.89  &       \\
J0838.8-2828  &  J083842.77-282830.9  &  19.66  &         &  18.49  &  19.02  &  18.01\\
J0841.3-3556  &  J084121.63-355505.9  &         &  17.20  &  17.57  &  16.72  &  16.64\\
J0844.9+6214  &  J084406.81+621458.6  &  19.39  &  16.38  &  18.75  &  16.62  &  15.85\\
J0858.3-4333  &  J085839.22-432642.7  &         &  20.10  &         &  20.12  &  17.83\\
J0900.9+6736  &  J090121.65+673955.8  &  19.09  &  18.52  &  19.61  &  18.25  &  18.13\\
J0955.0-3949  &  J095458.30-394655.0  &         &  16.66  &  18.15  &  17.21  &  17.11\\
J1013.6+3434  &  J101256.54+343648.8  &  20.22  &  18.60  &  20.44  &  17.99  &  17.39\\
J1016.1+5600  &  J101544.44+555100.7  &  19.69  &  19.42  &  20.61  &  19.35  &       \\
J1029.5-2022  &  J102946.66-201812.6  &  18.01  &  18.22  &  18.41  &  18.25  &  18.37\\
J1032.9-8401  &  J103015.35-840308.7  &         &  19.36  &  19.26  &  18.84  &  18.03\\
J1038.2-2423  &  J103754.92-242544.5  &  20.58  &  18.21  &  20.56  &  18.53  &  18.09\\
J1207.3-5055  &  J120746.43-505948.6  &         &         &  20.50  &  20.07  &       \\
              &  J120750.50-510314.9  &         &  19.12  &  19.71  &  20.27  &       \\
J1254.2-2203  &  J125422.47-220413.6  &         &  19.88  &  18.67  &  19.11  &  18.22\\
J1259.8-3749  &  J125949.80-374858.1  &         &  17.44  &  18.07  &  16.78  &  17.35\\
J1340.5-0412  &  J134042.02-041006.8  &  18.21  &  17.21  &  17.59  &  16.46  &  17.08\\
J1346.0-2605  &  J134621.08-255642.3  &  19.82  &  19.25  &  19.13  &  18.77  &  18.39\\
J1347.0-2956  &  J134706.89-295842.3  &  17.85  &  17.09  &  18.80  &  17.14  &  17.09\\
J1404.0-5244  &  J140313.11-524839.5  &         &  17.49  &  18.92  &  18.57  &       \\
J1517.2+3645  &  J151649.26+365022.9  &  20.90  &         &  21.49  &  20.07  &  19.16\\
J1612.0+1403  &  J161118.10+140328.9  &  18.39  &  18.39  &  19.06  &  19.16  &  18.43\\
J1614.8+4703  &  J161541.22+471111.8  &  17.55  &  16.03  &  16.90  &  15.39  &  15.51\\
              &  J161434.67+470420.1  &  15.62  &  15.76  &  16.28  &  16.13  &  15.14\\
              &  J161513.04+471355.2  &         &  20.02  &  21.44  &  20.29  &  19.09\\
              &  J161450.96+465953.7  &         &  19.19  &  21.66  &  19.60  &  19.10\\
J1622.8-0314  &  J162225.35-031439.6  &         &  19.85  &         &  19.41  &       \\
J1627.8+3219  &  J162800.40+322414.0  &  20.65  &  19.50  &  19.02  &  18.88  &  19.01\\
J1647.0+4351  &  J164619.95+435631.0  &  20.43  &  19.73  &  20.42  &  19.67  &       \\
J1730.6-0353  &  J173052.86-035247.1  &  18.31  &  17.40  &  19.30  &  17.33  &  16.73\\
J1745.6+0203  &  J174526.95+020532.7  &  18.71  &  17.12  &  18.06  &  17.28  &  17.16\\
              &  J174507.82+015442.5  &  19.21  &  16.40  &  18.11  &  16.30  &  15.98\\
J1759.2-3853  &  J175903.29-384739.5  &         &  17.95  &  18.95  &         &       \\
J1842.3+2740  &  J184201.25+274239.2  &  20.18  &  19.04  &  19.34  &  19.18  &  18.75\\
J1904.8-0705  &  J190444.57-070740.1  &         &  19.73  &  19.87  &  18.45  &       \\
J1924.9-1036  &  J192501.63-104316.3  &         &  18.63  &  19.42  &  18.04  &  17.75\\
J2004.6+7004  &  J200506.02+700439.3  &  20.73  &  19.25  &  19.24  &  18.65  &       \\
J2021.5+0632  &  J202155.45+062913.7  &  17.27  &  16.13  &  17.01  &  16.67  &  16.03\\
              &  J202154.66+062908.7  &  19.15  &  17.44  &  17.81  &  17.24  &  17.35\\
J2133.9+6645  &  J213349.21+664704.3  &         &         &         &  19.37  &  18.80\\
J2134.6-2130  &  J213430.18-213032.6  &  19.77  &  18.65  &  18.96  &  16.80  &  17.70\\
J2300.0-3553  &  J230010.16-360159.9  &         &  18.43  &  19.23  &  18.17  &  17.63\\
J2319.3-3830  &  J232000.11-383511.4  &         &  19.24  &  19.86  &  18.92  &  18.48\\
J2358.4-1811  &  J235828.61-181526.6  &  18.73  &  18.83  &  18.54  &  18.33  &  18.34\\
\noalign{\smallskip}
\hline
\end{tabular}\\
\label{tab:ugs_usno}
\end{table}

\begin{table}
\caption{Optical magnitudes of the USNO B1 catalog for the UGSs with $\gamma$-ray analysis flags.}
\tiny
\begin{tabular}{|llccccc|}
\hline
  2FGL   &   \wse\   &   B1     &   R1     &   B2     &   R2     &   I    \\ 
  name   &   name    &   mag    &   mag    &   mag    &   mag    &   mag   \\ 
\hline
\noalign{\smallskip}
J0233.9+6238c &  J023238.07+623651.9  &         &         &  20.95  &         &  18.70\\
              &  J023418.09+624207.8  &         &         &         &  19.34  &  17.89\\
J0341.8+3148c &  J034158.52+314855.7  &  18.76  &  14.98  &  17.64  &  14.86  &  13.62\\
              &  J034204.35+314711.4  &         &  19.84  &         &  19.86  &  17.33\\
J0440.5+2554c &  J043947.48+260140.5  &         &  19.09  &         &  19.11  &  15.70\\
J0620.8-2556  &  J062108.68-255757.9  &  19.78  &         &  19.94  &  19.22  &      \\
J0631.7+0428  &  J063104.12+042012.6  &         &  20.31  &         &  20.36  &       \\
J0637.0+0416c &  J063647.19+042058.7  &  20.25  &  17.34  &  20.42  &  17.61  &  16.47\\
              &  J063703.09+042146.1  &  18.95  &  16.95  &  19.88  &  17.23  &  16.04\\
              &  J063705.96+042537.2  &  21.00  &  18.79  &  20.37  &  17.45  &  16.87\\
              &  J063701.93+042037.2  &  21.00  &         &  18.57  &         &       \\
J0922.2-5214c &  J092154.24-521236.1  &         &         &  19.57  &  18.86  &  17.49\\
J1059.9-2051  &  J110025.72-205333.4  &  18.44  &  16.57  &  18.08  &  16.62  &  16.64\\
J1208.6-2257  &  J120816.33-224921.9  &  18.29  &  18.04  &  21.52  &  18.12  &  18.43\\
J1255.8-5828  &  J125459.44-582009.5  &         &  16.56  &  18.67  &  16.71  &  16.33\\
J1315.6-0730  &  J131543.62-073659.0  &  19.88  &  19.09  &  18.21  &  17.95  &  18.06\\
J1315.6-0730  &  J131552.98-073301.9  &  19.78  &  18.68  &  18.75  &  17.75  &  17.56\\
J1324.4-5411  &  J132415.49-541104.4  &         &  18.15  &         &  18.86  &       \\
J1345.8-3356  &  J134543.05-335643.3  &         &  17.98  &  19.58  &  18.65  &  18.12\\
J1407.4-2948  &  J140818.86-294203.2  &         &         &  20.80  &  19.36  &  18.75\\
J1624.2-2124  &  J162343.89-210707.0  &  19.70  &  18.57  &  19.35  &  18.74  &  18.51\\
J1835.4+1036  &  J183551.92+103056.8  &  18.37  &  16.97  &  17.97  &  16.73  &  16.30\\
J1835.4+1349  &  J183522.00+135733.9  &  19.69  &  18.29  &  19.36  &  18.15  &  17.24\\
              &  J183535.34+134848.8  &  19.57  &  17.15  &  19.01  &  16.66  &  16.84\\
J1837.9+3821  &  J183656.31+382232.8  &  18.37  &  18.01  &  19.53  &  18.45  &  18.34\\
              &  J183828.80+382704.3  &  20.75  &         &  20.96  &  20.60  &       \\
              &  J183837.16+381900.5  &  21.01  &  19.40  &  19.23  &  19.01  &  18.76\\
J1844.3+1548  &  J184425.36+154645.9  &  18.90  &  18.17  &  18.45  &  17.15  &  16.00\\
J1844.9-1116  &  J184456.29-111352.1  &         &  18.29  &         &         &  13.34\\
J1958.6+4020  &  J195842.28+401125.8  &  18.81  &         &  19.05  &         &       \\
J2124.0-1513  &  J212423.63-152558.2  &  20.18  &  19.41  &  20.33  &  18.87  &  18.59\\
J2128.7+5824  &  J212900.37+583128.0  &  20.28  &  18.42  &  19.96  &  18.28  &  17.71\\
\noalign{\smallskip}
\hline
\end{tabular}\\
\label{tab:ugf_usno}
\end{table}

\begin{table}
\caption{Optical magnitudes of the USNO B1 catalog for the AGUs (00h -- 12h).}
\tiny
\begin{tabular}{|llccccc|}
\hline
  2FGL  &  \wse\  &  B1    &  R1    &  B2    &  R2    &  I    \\ 
  name  &  name   &  mag   &  mag   &  mag   &  mag   &  mag   \\ 
\hline
\noalign{\smallskip}
J0009.1+5030  &  J000922.76+503028.8  &         &         &  19.74  &  19.35  &  17.32\\
J0018.8-8154  &  J001920.58-815251.3  &         &  15.86  &  16.62  &  16.13  &  15.33\\
J0022.2-1853  &  J002209.25-185334.7  &  19.05  &  18.63  &  18.07  &  16.95  &  17.29\\
J0022.3-5141  &  J002200.08-514024.2  &         &  15.65  &  17.38  &  15.94  &  16.65\\
J0045.5+1218  &  J004543.33+121712.0  &  18.22  &  17.40  &  18.75  &  16.91  &  15.66\\
J0051.4-6241  &  J005116.62-624204.3  &         &  16.95  &  16.83  &  16.69  &  15.78\\
J0059.7-5700  &  J005846.56-565911.4  &         &  17.45  &  17.39  &  16.95  &  17.07\\
J0110.3+6805  &  J011012.84+680541.1  &  18.66  &  16.51  &  18.14  &  16.33  &  15.37\\
J0134.4+2636  &  J013428.19+263843.0  &  16.87  &  16.74  &  17.09  &  15.91  &  15.55\\
J0156.4+3909  &  J015631.40+391430.9  &  17.99  &  18.05  &  18.71  &  19.03  &  18.25\\
J0156.5-2419  &  J015606.46-241754.3  &  21.01  &         &  20.86  &  21.40  &       \\
J0207.9-6832  &  J020750.91-683755.1  &         &  17.62  &  20.02  &  19.05  &  18.21\\
J0210.7-5102  &  J021046.19-510101.8  &         &  14.85  &  17.39  &  14.82  &  15.13\\
J0238.2-3905  &  J023749.42-390050.3  &         &  16.86  &  18.04  &  16.94  &  17.81\\
                    &  J023800.62-390504.6  &         &  17.52  &  17.63  &  16.59  &  16.96\\
J0248.6+8440  &  J024948.30+843556.9  &  19.50  &  18.43  &  19.42  &  17.62  &  17.01\\
J0253.4+3218  &  J025333.64+321720.8  &         &  19.83  &  20.39  &  19.69  &  19.46\\
J0309.3-0743  &  J030943.23-074427.4  &  18.33  &  15.96  &  18.16  &  16.01  &  16.71\\
J0332.5-1118  &  J033200.72-111456.1  &  17.78  &  17.08  &  18.85  &  17.75  &  18.30\\
                    &  J033223.25-111950.6  &  19.41  &  17.66  &  18.90  &  16.71  &  18.00\\
J0333.7+2918  &  J033349.00+291631.6  &  17.49  &  17.24  &  19.11  &  16.44  &  15.73\\
J0334.3+6538  &  J033356.74+653656.0  &         &  18.93  &  19.70  &  17.66  &  17.01\\
J0424.3-5332  &  J042347.22-533026.6  &         &  17.22  &  18.21  &  16.50  &  17.42\\
                    &  J042504.26-533158.3  &         &  15.54  &  17.41  &  16.14  &  16.18\\
J0433.9-5726  &  J043344.12-572613.3  &         &  17.73  &  18.90  &  17.78  &  18.67\\
J0438.8-4521  &  J043900.84-452222.6  &         &  18.37  &  20.85  &  19.48  &       \\
J0456.5+2658  &  J045617.36+270221.1  &  20.42  &         &  21.50  &         &  18.68\\
J0505.9+6116  &  J050558.78+611335.9  &         &  18.71  &  20.73  &  18.67  &  17.30\\
J0506.7-5435  &  J050657.80-543503.9  &         &  15.95  &  16.73  &  16.91  &  16.25\\
J0508.1-1936  &  J050805.75-194721.6  &  17.57  &  17.51  &  19.44  &  18.22  &  18.09\\
J0512.9+4040  &  J051252.53+404143.7  &  16.46  &  15.11  &  16.39  &  15.35  &  14.83\\
J0525.5-6011  &  J052537.74-601732.0  &         &  20.58  &         &  20.86  &       \\
J0532.0-4826  &  J053158.61-482736.0  &         &         &  20.61  &         &  18.98\\
J0537.7-5716  &  J053748.95-571830.2  &         &  17.23  &  18.07  &  17.17  &  17.83\\
J0609.4-0248  &  J060915.06-024754.6  &  18.05  &  17.47  &  18.23  &  16.92  &  16.47\\
J0621.9+3750  &  J062157.63+375057.0  &  20.30  &         &  20.29  &  19.96  &  18.63\\
J0644.2-6713  &  J064428.06-671257.3  &         &         &  20.47  &  20.87  &       \\
J0647.8-6102  &  J064806.55-610507.4  &         &  18.75  &  18.89  &  19.26  &  18.05\\
J0653.7+2818  &  J065344.26+281547.5  &  19.15  &  18.26  &  18.12  &  17.54  &  17.72\\
J0700.3+1710  &  J070001.50+170921.8  &  18.29  &  18.08  &  18.55  &  16.02  &  17.10\\
J0700.3+1710  &  J070046.29+171019.8  &  18.32  &  17.18  &  17.29  &  17.23  &  16.70\\
J0703.1-3912  &  J070312.64-391418.9  &         &  16.41  &  17.06  &  17.18  &  17.93\\
J0706.5+7741  &  J070651.32+774137.0  &  17.44  &  17.53  &  17.44  &  18.00  &  16.36\\
J0706.7-4845  &  J070549.12-483911.4  &         &  18.96  &  19.14  &  18.53  &  18.10\\
J0709.3-0256  &  J070945.05-025517.4  &         &         &  19.61  &  19.29  &       \\
J0726.0-0053  &  J072550.63-005456.4  &  17.63  &  16.58  &  17.41  &  15.82  &  15.91\\
J0734.2-7706  &  J073443.44-771113.4  &         &  19.76  &         &  20.77  &       \\
J0746.5-0713  &  J074627.48-070949.7  &  19.73  &  18.81  &  19.64  &  19.52  &  17.37\\
J0746.5-4758  &  J074642.30-475455.2  &         &  16.58  &  18.18  &  16.99  &  15.85\\
J0816.7-2420  &  J081639.46-242635.4  &  18.00  &         &  18.26  &  18.06  &  16.14\\
              &  J081640.41-242106.6  &  19.42  &         &  20.70  &         &       \\
J0823.0+4041  &  J082257.55+404149.8  &  19.00  &  18.7   &  19.34  &  19.47  &  18.16\\
J0844.8-5459  &  J084502.47-545808.5  &         &         &  19.35  &  16.80  &  17.65\\
J0849.9-3540  &  J084945.61-354101.2  &         &  19.03  &  20.27  &  18.45  &       \\
J0852.4-5756  &  J085238.73-575529.4  &         &  18.99  &  18.70  &  19.08  &  18.22\\
J0855.1-0712  &  J085435.20-071837.5  &  16.39  &  16.26  &  16.07  &  15.93  &  15.29\\
J0856.0+7136  &  J085654.85+714623.8  &  19.79  &  19.05  &         &  19.75  &  17.29\\
J0858.1-1952  &  J085805.36-195036.8  &  19.19  &  18.54  &  18.63  &  18.93  &  17.76\\
J0904.8-3513  &  J090423.42-351203.0  &         &  18.27  &  20.10  &  18.26  &       \\
J0906.2-0906  &  J090618.05-090544.9  &  19.06  &  19.13  &  19.07  &  18.59  &  18.04\\
J0919.3-2203  &  J092002.74-215835.0  &  18.55  &  17.99  &  19.26  &  19.24  &  18.12\\
J0940.8-6105  &  J094047.33-610728.5  &         &  16.51  &  18.03  &  16.68  &  16.29\\
J0941.9-0755  &  J094221.46-075953.1  &  19.82  &  18.84  &  17.66  &  17.91  &  18.08\\
J0946.9-2541  &  J094709.52-254100.0  &  16.68  &  16.80  &  18.16  &  16.71  &  16.62\\
J1016.2-0638  &  J101542.96-063055.1  &  19.24  &         &  19.94  &  19.70  &       \\
J1016.2-0638  &  J101626.98-063625.2  &  19.98  &         &  19.81  &  18.86  &  18.07\\
J1045.5-2931  &  J104540.62-292726.4  &  19.30  &  19.06  &  18.77  &  19.17  &  18.64\\
J1103.9-5356  &  J110352.22-535700.7  &         &  16.12  &  17.82  &  16.42  &  16.02\\
J1106.3-3643  &  J110624.04-364659.0  &         &  18.71  &  19.65  &  19.20  &  18.37\\
J1154.1-3242  &  J115406.16-324243.0  &         &  17.96  &  19.06  &  19.00  &  17.94\\
J1230.2-5258  &  J122939.88-530332.1  &         &  16.62  &  18.03  &  17.41  &  16.72\\
J1238.1-1953  &  J123824.40-195913.4  &  18.03  &  17.48  &  17.66  &  17.01  &  17.97\\
J1239.5+0728  &  J123924.58+073017.2  &  19.07  &  17.76  &  19.34  &  17.92  &  17.96\\
\noalign{\smallskip}
\hline
\end{tabular}\\
\label{tab:agu_usno1}
\end{table}

\begin{table}
\caption{Optical magnitudes of the USNO B1 catalog for the AGUs (12h -- 24h).}
\tiny
\begin{tabular}{|llccccc|}
\hline
  2FGL  &  \wse\  &  B1    &  R1    &  B2    &  R2    &  I    \\ 
  name  &  name   &  mag   &  mag   &  mag   &  mag   &  mag   \\ 
\hline
\noalign{\smallskip}
J1301.6+3331  &  J130147.03+332236.3  &  19.82  &  20.04  &  20.33  &  19.38  &  18.92\\
J1303.8-5537  &  J130349.23-554031.6  &         &  16.82  &         &  17.92  &  17.63\\
J1304.1-2415  &  J130416.70-241216.6  &  17.10  &  16.56  &  19.94  &  18.16  &  16.60\\
J1304.3-4353  &  J130421.01-435310.2  &         &  15.75  &  17.48  &  16.10  &  15.48\\
J1307.5-4300  &  J130737.98-425938.9  &         &  15.75  &  16.26  &  15.58  &  14.98\\
J1307.6-6704  &  J130817.51-670705.8  &         &         &  20.40  &  17.68  &       \\
J1329.2-5608  &  J132901.16-560802.5  &         &         &  17.38  &         &  17.11\\
J1330.1-7002  &  J133011.34-700312.7  &         &         &  16.45  &  17.37  &       \\
J1351.3-2909  &  J135146.85-291217.4  &         &         &  20.22  &  19.44  &       \\
J1406.2-2510  &  J140609.60-250809.2  &  16.79  &  16.25  &  16.20  &  16.51  &  15.95\\
J1416.3-2415  &  J141642.23-241021.2  &  18.39  &  18.16  &  19.02  &  18.83  &  18.02\\
J1419.4-0835  &  J141922.56-083831.9  &  20.95  &  19.28  &  18.75  &  19.55  &  18.60\\
J1514.6-4751  &  J151440.03-474829.7  &         &  17.61  &         &  18.44  &  17.17\\
J1518.2-2733  &  J151803.60-273131.0  &  18.40  &  15.97  &  16.62  &  14.21  &  15.18\\
J1553.2-2424  &  J155331.62-242206.0  &  19.67  &  18.74  &  20.43  &  18.70  &  16.94\\
J1558.3+8513  &  J160031.76+850949.2  &  19.66  &  18.78  &  19.50  &  18.79  &  17.90\\
J1604.5-4442  &  J160431.03-444131.9  &         &  17.50  &  20.00  &         &       \\
J1626.0-7636  &  J162638.17-763855.4  &         &  14.89  &  16.12  &  14.94  &  15.05\\
J1725.1-7714  &  J172350.86-771350.3  &         &  19.12  &  19.71  &  18.94  &  18.31\\
J1811.0+5340  &  J181037.99+533501.5  &  18.88  &  18.77  &  19.34  &  18.45  &  16.71\\
J1815.6-6407  &  J181425.96-641008.8  &         &  18.87  &  18.96  &  19.02  &       \\
J1816.7-4942  &  J181655.99-494344.7  &         &  18.23  &  17.88  &  18.31  &  17.92\\
J1818.7+2138  &  J181905.22+213234.0  &  17.64  &  16.75  &  17.52  &  17.22  &  16.49\\
J1820.6+3625  &  J182023.61+362914.4  &  17.93  &  18.20  &         &  17.94  &  17.43\\
J1825.1-5231  &  J182513.79-523058.1  &         &  19.05  &  18.80  &  18.83  &  16.74\\
J1830.0+1325  &  J183000.76+132414.4  &  19.91  &  17.77  &  18.57  &  17.66  &  17.51\\
J1830.2-4441  &  J183000.86-444111.4  &         &  16.54  &  18.17  &  17.47  &  16.88\\
J1844.7+5716  &  J184450.96+570938.6  &  17.69  &  17.79  &  18.52  &  17.49  &  17.55\\
J1936.9+8402  &  J193930.23+835925.8  &  19.59  &  18.55  &  18.89  &  19.11  &  18.45\\
J1940.8-6213  &  J194121.76-621120.8  &         &         &  21.07  &  20.04  &  18.53\\
J1942.8+1033  &  J194247.48+103327.2  &  18.59  &  16.82  &  16.69  &  15.37  &  15.24\\
J1959.9-4727  &  J195945.66-472519.2  &         &  16.61  &  16.66  &  16.70  &  16.48\\
J2103.6-6236  &  J210338.38-623225.8  &         &  16.00  &  17.83  &  16.24  &  16.13\\
J2250.2-4205  &  J225014.94-420218.6  &         &  19.52  &  20.53  &  19.44  &  18.46\\
              &  J225022.20-420613.2  &         &  16.21  &  17.41  &  17.23  &  16.49\\
J2317.3-4534  &  J231731.97-453359.6  &         &  17.36  &  18.24  &  18.80  &  17.44\\
J2323.0-4918  &  J232255.30-491942.0  &         &  20.55  &  20.19  &  20.13  &  17.67\\
J2324.6+0801  &  J232445.31+080206.3  &  18.75  &  17.97  &  18.63  &  17.37  &  17.92\\
J2325.4+1650  &  J232526.62+164941.1  &  20.45  &         &         &  20.50  &       \\
              &  J232538.11+164642.8  &  18.56  &  18.29  &  17.37  &  17.09  &  17.27\\
\noalign{\smallskip}
\hline
\end{tabular}\\
\label{tab:agu_usno2}
\end{table}

\end{document}